**Rajesh Singh, Pankaj Chauhan,
Nirmala Sawan, Florentin Smarandache**

# Auxiliary Information and *A Priori* Values in Construction of Improved Estimators

| *Estimators* | *PRE* $(., s_y^2)$ | |
|---|---|---|
| | *Population I* | *Population II* |
| $s_y^2$ | 100 | 100 |
| *t₁* | 223.14 | 228.70 |
| *t₂* | 235.19 | 228.76 |
| *tᵣ (optimum)* | 305.66 | 232.90 |
| *tₚ (optimum)* | 305.66 | 232.90 |

*PRE of different estimators of $S_y^2$ with respect to $s_y^2$*

**2007**

# Auxiliary Information and a priori Values in Construction of Improved Estimators

Rajesh Singh, Pankaj Chauhan,
Nirmala Sawan
School of Statistics, DAVV,
Indore (M. P.), India

Florentin Smarandache
Department of Mathematics,
University of New Mexico, Gallup, USA

2007

# Contents

# Preface

This volume is a collection of six papers on the use of auxiliary information and *a priori* values in construction of improved estimators. The work included here will be of immense application for researchers and students who employ auxiliary information in any form.

Below we discuss each paper:

1. **Ratio estimators in simple random sampling using information on auxiliary attribute.**

    Prior knowledge about population mean along with coefficient of variation of the population of an auxiliary variable is known to be very useful particularly when the ratio, product and regression estimators are used for estimation of population mean of a variable of interest. However, the fact that the known population proportion of an attribute also provides similar type of information has not drawn as much attention. In fact, such prior knowledge can also be very useful when a relation between the presence (or absence) of an attribute and the value of a variable, known as point biserial correlation, is observed. Taking into consideration the point biserial correlation between a variable and an attribute, Naik and Gupta (1996) defined ratio, product and regression estimators of population mean when the prior information of population proportion of units, possessing the same attribute is available. In the present paper, some ratio estimators for estimating the population mean of the variable under study, which make use of information regarding the population proportion possessing certain attribute are proposed. The expressions of bias and mean squared error (MSE) have been obtained. The results obtained have been illustrated numerically by taking some empirical populations considered in the literature.

2. **Ratio-Product type exponential estimator for estimating finite population mean using information on auxiliary attribute.**

It is common practice to use arithmetic mean while constructing estimators for estimating population mean. Mohanty and Pattanaik (1984) used geometric mean and harmonic mean, while constructing estimators for estimating population mean, using multi-auxiliary variables. They have shown that in case of multi-auxiliary variables, estimates based on geometric mean and harmonic means are less biased than Olkin's (1958) estimate based on arithmetic mean under certain conditions usually satisfied in practice. For improving the precision in estimating the unknown mean $\overline{Y}$ of a finite population by using the auxiliary variable x, which may be positively or negatively correlated with y with known $\overline{X}$; the single supplementary variable is used by Bahl and Tuteja (1991) for the exponential ratio and product type estimators. For estimating the population mean $\overline{Y}$ of the study variable y, following Bahl and Tuteja (1991), a ratio-product type exponential estimator has been proposed by using the known information of population proportion possessing an attribute (highly correlated with y) in simple random sampling. The proposed estimator has an improvement over mean per unit estimator, ratio and product type exponential estimators as well as Naik and Gupta (1996) estimators. The results have also been extended to the case of two-phase sampling.

## 3. Improvement in estimating the population mean using exponential estimator in simple random sampling.

Using known values of certain population parameter(s) including coefficient of variation, coefficient of variation, coefficient of kurtosis, correlation coefficient, several authors have suggested modified ratio estimators for estimating population mean $\overline{Y}$. In this paper, under simple random sampling without replacement (SRSWOR), authors have suggested improved exponential ratio-type estimator for estimating population mean using some known values of population parameter(s). An empirical study is carried out to show the properties of the proposed estimator.

## 4. Almost unbiased exponential estimator for the finite population mean.

Usual ratio and product estimators and also exponential ratio and product type estimators suggested by Bahl and Tuteja (1991) are biased. Biasedness of an estimator is disadvantageous in some applications. This encouraged many researchers including

Hartley and Ross (1954) and Singh and Singh (1992) to construct either estimator with reduced bias known as almost unbiased estimator or completely unbiased estimator. In this paper we have proposed an almost unbiased ratio and product type exponential estimator for the finite population mean $\overline{Y}$. It has been shown that Bahl and Tuteja (1991) ratio and product type exponential estimators are particular members of the proposed estimator. Empirical study is carried to demonstrate the superiority of the proposed estimator.

## 5. Almost unbiased ratio and product type estimator of finite population variance using the knowledge of kurtosis of an auxiliary variable in sample surveys.

In manufacturing industries and pharmaceutical laboratories sometimes researchers are interested in the variation of their product or yields. Using the knowledge of kurtosis of auxiliary variable Upadhyaya and Singh (1999) have suggested an estimator for population variance. In this paper following the approach of Singh and Singh (1993), we have suggested almost unbiased ratio and product type estimator for population variance.

## 6. A general family of estimators for estimating population variance using known value of some population parameter(s).

In this paper, a general family of estimators for estimating the population variance of the variable under study; using known values of certain population parameter(s) is proposed. It has been shown that some existing estimators in literature are particular member of the proposed class. An empirical study is caring out to illustrate the performance of the constructed estimator over other.

<div style="text-align: right;">The Authors</div>

# Ratio Estimators in Simple Random Sampling Using Information on Auxiliary Attribute

Rajesh Singh, Pankaj Chauhan, Nirmala Sawan,

School of Statistics, DAVV, Indore (M.P.), India

(rsinghstat@yahoo.com)

Florentin Smarandache

Chair of Department of Mathematics, University of New Mexico, Gallup, USA

(smarand@unm.edu)

**Abstract**

Some ratio estimators for estimating the population mean of the variable under study, which make use of information regarding the population proportion possessing certain attribute, are proposed. Under simple random sampling without replacement (SRSWOR) scheme, the expressions of bias and mean-squared error (MSE) up to the first order of approximation are derived. The results obtained have been illustrated numerically by taking some empirical population considered in the literature.

## 1. Introduction

The use of auxiliary information can increase the precision of an estimator when study variable y is highly correlated with auxiliary variable x. There exist situations when information is available in the form of attribute $\phi$, which is highly correlated with y. For example

    a) Sex and height of the persons,

    b) Amount of milk produced and a particular breed of the cow,

    c) Amount of yield of wheat crop and a particular variety of wheat etc. (see Jhajj et al., [1]).

Consider a sample of size n drawn by SRSWOR from a population of size N. Let $y_i$ and $\phi_i$ denote the observations on variable y and $\phi$ respectively for $i^{th}$ unit $(i=1,2,....N)$. Suppose there is a complete dichotomy in the population with respect to the presence or absence of an attribute, say $\phi$, and it is assumed that attribute $\phi$ takes only the two values 0 and 1 according as

$\phi_i$ = 1, if ith unit of the population possesses attribute $\phi$

    = 0, otherwise.

Let $A = \sum_{i=1}^{N} \phi_i$ and $a = \sum_{i=1}^{n} \phi_i$ denote the total number of units in the population and sample respectively possessing attribute $\phi$. Let $P = \dfrac{A}{N}$ and $p = \dfrac{a}{n}$ denote the proportion of units in the population and sample respectively possessing attribute $\phi$.

Taking into consideration the point biserial correlation between a variable and an attribute, Naik and Gupta (1996) defined ratio estimator of population mean when the prior information of population proportion of units, possessing the same attribute is available, as follows:

$$t_{NG} = \bar{y}\left(\frac{P}{p}\right) \qquad (1.1)$$

here $\bar{y}$ is the sample mean of variable of interest. The MSE of $t_{NG}$ up to the first order of approximation is

$$MSE(t_{NG}) = \left(\frac{1-f}{n}\right)\left[S_y^2 + R_1^2 S_\phi^2 - 2R_1 S_{y\phi}\right] \qquad (1.2)$$

where $f = \dfrac{n}{N}$, $R_1 = \dfrac{\bar{Y}}{P}$, $S_y^2 = \dfrac{1}{N-1}\sum_{i=1}^{N}(y_i - \bar{Y})^2$, $S_\phi^2 = \dfrac{1}{N-1}\sum_{i=1}^{N}(\phi_i - P)^2$,

$S_{y\phi} = \dfrac{1}{N-1}\sum_{i=1}^{N}(\phi_i - P)(y_i - \bar{Y})$.

In the present paper, some ratio estimators for estimating the population mean of the variable under study, which make use of information regarding the population proportion possessing certain attribute, are proposed. The expressions of bias and MSE have been obtained. The numerical illustrations have also been done by taking some empirical populations considered in the literature.

## 2. The suggested estimator

Following Ray and Singh (1981), we propose the following estimator

$$t_1 = \frac{\bar{y} + b_\phi(P-p)}{p}P = R^*P \qquad (2.1)$$

where $b_\phi = \dfrac{s_{y\phi}}{s_\phi^2}$, $R^* = \dfrac{\bar{y} + b_\phi(P-p)}{p}$, $s_\phi^2 = \left(\dfrac{1}{n-1}\right)\sum_{i=1}^{n}(\phi_i - p)^2$ and

$s_{y\phi} = \left(\dfrac{1}{n-1}\right)\sum_{i=1}^{n}(\phi_i - p)(y_i - \bar{Y})$.

**Remark 1**: When we put $b_\phi = 0$ in (2.1), the proposed estimator turns to the Naik and Gupta (1996) ratio estimator $t_{NG}$ given in (1.1).

MSE of this estimator can be found by using Taylor series expansion given by

$$f(p,\bar{y}) \cong f(P,\bar{Y}) + \frac{\partial f(c,d)}{\partial c}\bigg|_{P,\bar{Y}}(p-P) + \frac{\partial f(c,d)}{\partial c}\bigg|_{P,\bar{Y}}(\bar{y}-\bar{Y}) \qquad (2.2)$$

where $f(p,\bar{y}) = R^*$ and $f(P,\bar{Y}) = R_1$.

Expression (2.2) can be applied to the proposed estimator in order to obtain MSE equation as follows:

$$R^* - R_1 \cong \frac{\partial\left((\bar{y}+b_\phi(P-p))/p\right)}{\partial p}\bigg|_{P,\bar{Y}}(p-P) + \frac{\partial\left((\bar{y}+b_\phi(P-p))/p\right)}{\partial \bar{y}}\bigg|_{P,\bar{Y}}(\bar{y}-\bar{Y})$$

$$\cong -\left(\frac{\bar{y}}{p^2} + \frac{b_\phi P}{p^2}\right)\bigg|_{P,\bar{Y}}(p-P) + \frac{1}{p}\bigg|_{P,\bar{Y}}(\bar{y}-\bar{Y})$$

$$E(R^* - R_1)^2 \cong \frac{(\bar{Y}+B_\phi P)^2}{P^4}V(p) - \frac{2(\bar{Y}+B_\phi P)}{P^3}Cov(p,\bar{y}) + \frac{1}{P^2}V(\bar{y})$$

$$\cong \frac{1}{P^2}\left\{\frac{(\bar{Y}+B_\phi P)^2}{P^2}V(p) - \frac{2(\bar{Y}+B_\phi P)}{P}Cov(p,\bar{y}) + V(\bar{y})\right\} \qquad (2.3)$$

where $B_\phi = \dfrac{S_{\phi y}}{S_\phi^2} = \dfrac{\rho_{pb} S_y}{S_\phi}$.

$\rho_{pb} = \dfrac{S_{y\phi}}{S_y S_\phi}$, is the point biserial correlation coefficient.

Now,

$$MSE(t_1) = P^2 E(R_1 - R_\phi)^2$$

$$\cong \frac{(\bar{Y}+B_\phi P)^2}{P^2}V(p) - \frac{2(\bar{Y}+B_\phi P)}{P}Cov(p,\bar{y}) + V(\bar{y}) \qquad (2.4)$$

Simplifying (2.4), we get MSE of $t_1$ as

$$\text{MSE}(t_1) \cong \left(\frac{1-f}{n}\right)\left[R_1^2 S_\phi^2 + S_y^2\left(1 - \rho_{pb}^2\right)\right] \tag{2.5}$$

Several authors have used prior value of certain population parameters (s) to find more precise estimates. Searls (1964) used Coefficient of Variation (CV) of study character at estimation stage. In practice this CV is seldom known. Motivated by Searls (1964) work, Sen (1978), Sisodiya and Dwivedi (1981), and Upadhyaya and Singh (1984) used the known CV of the auxiliary character for estimating population mean of a study character in ratio method of estimation. The use of prior value of Coefficient of Kurtosis in estimating the population variance of study character y was first made by Singh et al. (1973). Later, used by and Searls and Intarapanich (1990), Upadhyaya and Singh (1999), Singh (2003) and Singh et al. (2004) in the estimation of population mean of study character. Recently Singh and Tailor (2003) proposed a modified ratio estimator by using the known value of correlation coefficient.

In next section, we propose some ratio estimators for estimating the population mean of the variable under study using known parameters of the attribute $\phi$ such as coefficient of variation $C_p$, Kurtosis $(\beta_2(\phi))$ and point biserial correlation coefficient $\rho_{pb}$.

### 3. Suggested Estimators

We suggest following estimator

$$t = \frac{\bar{y} + b_\phi(P - p)}{(m_1 p + m_2)}(m_1 P + m_2) \tag{3.1}$$

where $m_1(\neq 0)$, $m_2$ are either real number or the functions of the known parameters of the attribute such as $C_p$, $(\beta_2(\phi))$ and $\rho_{pb}$.

The following scheme presents some of the important estimators of the population mean, which can be obtained by suitable choice of constants $m_1$ and $m_2$:

| Estimator | Values of | |
|---|---|---|
| | $m_1$ | $m_2$ |
| $t_1 = \dfrac{\bar{y} + b_\phi(P-p)}{p} P$ | 1 | 1 |
| $t_2 = \dfrac{\bar{y} + b_\phi(P-p)}{(p + \beta_2(\phi))}[P + \beta_2(\phi)]$ | 1 | $\beta_2(\phi)$ |
| $t_3 = \dfrac{\bar{y} + b_\phi(P-p)}{(p + C_p)}[P + C_p]$ | 1 | $C_p$ |
| $t_4 = \dfrac{\bar{y} + b_\phi(P-p)}{(p + \rho_{pb})}[P + \rho_{pb}]$ | 1 | $\rho_{pb}$ |
| $t_5 = \dfrac{\bar{y} + b_\phi(P-p)}{(p\beta_2(\phi) + C_p)}[P\beta_2(\phi) + C_p]$ | $\beta_2(\phi)$ | $C_p$ |
| $t_6 = \dfrac{\bar{y} + b_\phi(P-p)}{(pC_p + \beta_2(\phi))}[PC_p + \beta_2(\phi)]$ | $C_p$ | $\beta_2(\phi)$ |
| $t_7 = \dfrac{\bar{y} + b_\phi(P-p)}{(pC_p + \rho_{pb})}[PC_p + \rho_{pb}]$ | $C_p$ | $\rho_{pb}$ |
| $t_8 = \dfrac{\bar{y} + b_\phi(P-p)}{(p\rho_{pb} + C_p)}[P\rho_{pb} + C_p]$ | $\rho_{pb}$ | $C_p$ |
| $t_9 = \dfrac{\bar{y} + b_\phi(P-p)}{(p\beta_2(\phi) + \rho_{pb})}[P\beta_2(\phi) + \rho_{pb}]$ | $\beta_2(\phi)$ | $\rho_{pb}$ |
| $t_{10} = \dfrac{\bar{y} + b_\phi(P-p)}{(p\rho_{pb} + \beta_2(\phi))}[P\rho_{pb} + \beta_2(\phi)]$ | $\rho_{pb}$ | $\beta_2(\phi)$ |

Following the approach of section 2, we obtain the MSE expression for these proposed estimators as –

$$\text{MSE}(t_i) \cong \left(\frac{1-f}{n}\right)\left[R_i S_\phi^2 + S_y^2(1-\rho_{pb}^2)\right], \qquad (i=1,2,3,\ldots,10) \qquad (3.2)$$

where $R_1 = \dfrac{\overline{Y}}{P}$, $R_2 = \dfrac{\overline{Y}}{P+\beta_2(\phi)}$, $R_3 = \dfrac{\overline{Y}}{P+C_p}$, $R_4 = \dfrac{\overline{Y}}{P+\rho_{pb}}$,

$R_5 = \dfrac{\overline{Y}\beta_2(\phi)}{P\beta_2(\phi)+C_p}$, $R_6 = \dfrac{\overline{Y}C_p}{PC_p+\beta_2(\phi)}$, $R_7 = \dfrac{\overline{Y}C_p}{PC_p+\rho_{pb}}$, $R_8 = \dfrac{\overline{Y}\rho_{pb}}{P\rho_{pb}+C_p}$,

$R_9 = \dfrac{\overline{Y}\beta_2(\phi)}{P\beta_2(\phi)+\rho_{pb}}$, $R_{10} = \dfrac{\overline{Y}\rho_{pb}}{P\rho_{pb}+\beta_2(\phi)}$.

## 4. Efficiency comparisons

It is well known that under simple random sampling without replacement (SRSWOR) the variance of the sample mean is

$$V(\overline{y}) = \left(\frac{1-f}{n}\right)S_y^2$$

(4.1)

From (4.1) and (3.2), we have

$$V(\overline{y}) - \text{MSE}(t_i) \geq 0, \qquad i=1,2,\ldots,10$$

$$\Rightarrow \rho_{pb}^2 > \frac{S_\phi^2}{S_y^2}R_i^2$$

(4.2)

When this condition is satisfied, proposed estimators are more efficient than the sample mean.

Now, we compare the MSE of the proposed estimators with the MSE of Naik and Gupta [2] estimator $t_{NG}$. From (3.2) and (1.1) we have

$$\text{MSE}(t_{NG}) - \text{MSE}(t_i) \geq 0, \quad (i = 1,2,\ldots,10)$$

$$\Rightarrow \rho_{pb}^2 \geq \frac{S_\phi^2}{S_y^2}\left[R_i^2 - R_\phi^2 + 2R_\phi K_{yp}\right] \tag{4.3}$$

where $K_{yp} = \rho_{yp}\dfrac{C_y}{C_p}$.

## 5. Empirical Study

The data for the empirical study is taken from natural population data set considered by Sukhatme and Sukhatme [12]:

y = Number of villages in the circles and

$\phi$ = A circle consisting more than five villages

$N = 89$, $\overline{Y} = 3.36$, $P = 0.1236$, $\rho_{pb} = 0.766$, $C_y = 0.604$, $C_p = 2.19$, $\beta_2(\phi) = 6.23181$.

In the below table 5.1 percent relative efficiencies (PRE) of various estimators are computed with respect to $\overline{y}$.

**Table 5.1: PRE of different estimators of $\overline{Y}$ with respect to $\overline{y}$.**

| Estimator | PRE $(., \overline{y})$ |
|---|---|
| $\overline{y}$ | 100 |
| $t_{NG}$ | 11.61 |
| $t_1$ | 7.36 |
| $t_2$ | 236.55 |
| $t_3$ | 227.69 |
| $t_4$ | 208.09 |
| $t_5$ | 185.42 |
| $t_6$ | 230.72 |
| $t_7$ | 185.27 |
| $t_8$ | 230.77 |
| $t_9$ | 152.37 |
| $t_{10}$ | 237.81 |

From table 5.1, we observe that the proposed estimators $t_i (i = 2,....,10)$ which uses some known values of population proportion performs better than the usual sample mean $\overline{y}$ and Naik and Gupta [2] estimator $t_{NG}$.

**Conclusion:**

We have suggested some ratio estimators for estimating $\overline{Y}$ which uses some known value of population proportion. For practical purposes the choice of the estimator depends upon the availability of the population parameters.

**References**

Jhajj, H. S., Sharma, M. K. and Grover, L. K., A family of estimators of population mean using information on auxiliary attribute. *Pakistan Journal of Statistics*, 22 (1), 43-50 (2006).

Naik, V. D. and Gupta, P. C., A note on estimation of mean with known population proportion of an auxiliary character. *Journal of the Indian Society of Agricultural Statistics*, 48 (2), 151-158 (1996).

Ray, S. K. and Singh, R. K., Difference-cum-ratio type estimators. *Journal of the Indian Statistical Association*, 19, 147-151 (1981).

Searls, D. T., The utilization of known coefficient of variation in the estimation procedure. *Journal of the American Statistical Association*, **59**, 1125-1126 (1964).

Searls, D. T. and Intarapanich, P., A note on an estimator for the variance that utilizes the kurtosis. *The American Statistician*, **44**, 295-296 (1990).

Sen, A. R., Estimation of the population mean when the coefficient of variation is known. *Communications in Statistics Theory and Methods A*, **7**, 657-672 (1978).

Singh, G. N., On the improvement of product method of estimation in sample surveys. *Journal of the Indian Society of Agricultural Statistics*, **56** (3), 267-275 (2003).

Singh H. P. and Tailor, R., Use of known correlation coefficient in estimating the finite population mean. *Statistics in Transition*, 6, 555-560 (2003).

Singh H. P., Tailor, R., Tailor, R. and Kakran, M. S., An improved estimator of population mean using power transformation. *Journal of the Indian Society of Agricultural Statistics*, 58 (2), 223-230 (2004).

Singh, J., Pandey, B. N. and Hirano, K., On the utilization of a known coefficient of kurtosis in the estimation procedure of variance. *Annals of the Institute of Statistical Mathematics*, **25**, 51-55 (1973).

Sisodia, B. V. S. and Dwivedi, V. K., A modified ratio estimator using coefficient of variation of auxiliary variable. *Journal of the Indian Society of Agricultural Statistics*, **33** (2), 13-18 (1981).

Sukhatme, P. V. and Sukhatme, B. V., *Sampling Theory of Surveys with Applications*. Iowa State University Press, Ames IOWA, 1970.

Upadhyaya, L. N. and Singh, H. P., On the estimation of the population mean with known coefficient of variation. *Biometrical Journal*, **26**, 915-922 (1984).

Upadhyaya, L. N. and Singh, H. P., Use of transformed auxiliary variable in estimating the finite population mean. *Biometrical Journal*, **41**, 627-636 (1999).

# Ratio-Product Type Exponential Estimator for Estimating Finite Population Mean Using Information on Auxiliary Attribute

Rajesh Singh, Pankaj Chauhan, Nirmala Sawan

School of Statistics, DAVV, Indore (M.P.), India

(rsinghstat@yahoo.com)

Florentin Smarandache

Chair of Department of Mathematics, University of New Mexico, Gallup, USA

(smarand@unm.edu)

**Abstract**

In practice, the information regarding the population proportion possessing certain attribute is easily available, see Jhajj et.al. (2006). For estimating the population mean $\overline{Y}$ of the study variable y, following Bahl and Tuteja (1991), a ratio-product type exponential estimator has been proposed by using the known information of population proportion possessing an attribute (highly correlated with y) in simple random sampling. The expressions for the bias and the mean-squared error (MSE) of the estimator and its minimum value have been obtained. The proposed estimator has an improvement over mean per unit estimator, ratio and product type exponential estimators as well as Naik and Gupta (1996) estimators. The results have also been extended to the case of two phase sampling. The results obtained have been illustrated numerically by taking some empirical populations considered in the literature.

**Keywords:** Proportion, bias, mean-squared error, two phase sampling.

## 1. Introduction

In survey sampling, the use of auxiliary information can increase the precision of an estimator when study variable y is highly correlated with the auxiliary variable x. but in several practical situations, instead of existence of auxiliary variables there exists some auxiliary attributes, which are highly correlated with study variable y, such as

(i) Amount of milk produced and a particular breed of cow. (ii) Yield of wheat crop and a particular variety of wheat etc. (see Shabbir and Gupta (2006)).

In such situations, taking the advantage of point biserial correlation between the study variable and the auxiliary attribute, the estimators of parameters of interest can be constructed by using prior knowledge of the parameters of auxiliary attribute.

Consider a sample of size n drawn by simple random sampling without replacement (SRSWOR) from a population of size N. let $y_i$ and $\phi_i$ denote the observations on variable y and $\phi$ respectively for the $i^{th}$ unit ($i = 1,2,...,N$). We note that $\phi_i = 1$, if $i^{th}$ unit of population possesses attribute $\phi$ and $\phi_i = 0$, otherwise. Let $A = \sum_{i=1}^{N} \phi_i$ and $a = \sum_{i=1}^{n} \phi_i$ denote the total number of units in the population and sample respectively possessing attribute $\phi$. Let $P = \frac{A}{N}$ and $p = \frac{a}{n}$ denote the proportion of units in the population and sample respectively possessing attribute $\phi$.

In order to have an estimate of the population mean $\overline{Y}$ of the study variable y, assuming the knowledge of the population proportion P, Naik and Gupta (1996) defined ratio and product estimators of population when the prior information of population proportion of units, possessing the same attribute is available. Naik and Gupta (1996) proposed following estimators:

$$t_1 = \bar{y}\left(\frac{P}{p}\right) \tag{1.1}$$

$$t_2 = \bar{y}\left(\frac{p}{P}\right) \tag{1.2}$$

The MSE of $t_1$ and $t_2$ up to the first order of approximation are

$$MSE(t_1) = f_1 \bar{Y}^2 \left[C_y^2 + C_p^2(1 - 2K_p)\right] \tag{1.3}$$

$$MSE(t_2) = f_1 \bar{Y}^2 \left[C_y^2 + C_p^2(1 + 2K_p)\right] \tag{1.4}$$

where $C_y^2 = \dfrac{S_y^2}{\bar{Y}^2}, C_p^2 = \dfrac{S_\phi^2}{P^2}, f_1 = \dfrac{1}{n} - \dfrac{1}{N}, K_p = \rho_{pb}\dfrac{C_y}{C_p}, S_y^2 = \dfrac{1}{N-1}\sum_{i=1}^{N}(y_i - \bar{Y})^2,$

$S_\phi^2 = \dfrac{1}{N-1}\sum_{i=1}^{N}(\phi_i - P)^2, S_{y\phi} = \dfrac{1}{N-1}\left(\sum_{i=1}^{N} y_i \phi_i - NP\bar{Y}\right)$ and

$\rho_{pb} = \dfrac{S_{y\phi}}{S_y S_\phi}$ is the point biserial correlation coefficient.

Following Bahl and Tuteja (1991), we propose the following ratio and product exponential estimators

$$t_3 = \bar{y}\exp\left(\frac{P-p}{P+p}\right) \tag{1.5}$$

$$t_4 = \bar{y}\exp\left(\frac{p-P}{p+P}\right) \tag{1.6}$$

## 2. Bias and MSE of $t_3$ and $t_4$

To obtain the bias and MSE of $t_3$ to the first degree of approximation, we define

$e_y = \dfrac{(\bar{y} - \bar{Y})}{\bar{Y}}, e_\phi = \dfrac{(p - P)}{P}$, therefore E ($e_i$) = 0. i = (y, φ),

$E(e_y^2) = f_1 C_y^2, \ E(e_\phi^2) = f_1 C_p^2, E(e_y e_\phi) = f_1 \rho_{pb} C_y C_p.$

Expressing (1.5) in terms of e's, we have

$$t_3 = \overline{Y}(1+e_y)\exp\left[\frac{P-P(1+e_\phi)}{P+P(1+e_\phi)}\right]$$

$$= \overline{Y}(1+e_y)\exp\left[\frac{-e_\phi}{(2+e_\phi)}\right] \qquad (2.1)$$

Expanding the right hand side of (2.1) and retaining terms up to second powers of e's, we have

$$t_3 = \overline{Y}\left[1+e_y - \frac{e_\phi}{2} + \frac{e_\phi^2}{8} - \frac{e_y e_\phi}{2}\right] \qquad (2.2)$$

Taking expectations of both sides of (2.2) and then subtracting $\overline{Y}$ from both sides, we get the bias of the estimator $t_3$ up to the first order of approximation, as

$$B(t_3) = f_1 \overline{Y} \frac{C_p^2}{2}\left(\frac{1}{4} - K_p\right) \qquad (2.3)$$

From (2.2), we have

$$(t_3 - \overline{Y}) \cong \overline{Y}\left[e_y - \frac{e_\phi}{2}\right] \qquad (2.4)$$

Squaring both sides of (2.4) and then taking expectations we get MSE of the estimator $t_3$, up to the first order of approximation as

$$MSE(t_3) = f_1 \overline{Y}^2 \left[C_y^2 + C_p^2(\frac{1}{4} - K_p)\right] \qquad (2.5)$$

To obtain the bias and MSE of $t_4$ to the first degree of approximation, we express (1.6) in terms of e's

$$t_4 = \overline{Y}(1+e_y)\exp\left[\frac{P(1+e_\phi)-P}{P(1+e_\phi)+P}\right] \qquad (2.6)$$

and following the above procedure, we get the bias and MSE of $t_4$ as follows

$$B(t_4) = f_1\overline{Y}\frac{C_p^2}{2}\left(\frac{1}{4}+K_p\right) \qquad (2.7)$$

$$MSE(t_4) = f_1\overline{Y}^2\left[C_y^2 + C_p^2(\frac{1}{4}+K_p)\right] \qquad (2.8)$$

### 3. Proposed class of estimators

It has been theoretically established that, in general, the linear regression estimator is more efficient than the ratio (product) estimator except when the regression line of y on x passes through the neighborhood of the origin, in which case the efficiencies of these estimators are almost equal. Also in many practical situations the regression line does not pass through the neighborhood of the origin. In these situations, the ratio estimator does not perform as good as the linear regression estimator. The ratio estimator does not perform well as the linear regression estimator does.

Following Singh and Espejo (2003), we propose following class of ratio-product type exponential estimators:

$$t_5 = \overline{y}\left[\alpha\exp\left(\frac{P-p}{P+p}\right) + (1-\alpha)\exp\left(\frac{p-P}{p+P}\right)\right] \qquad (3.1)$$

where $\alpha$ is a real constant to be determined such that the MSE of $t_5$ is minimum.

For $\alpha=1$, $t_5$ reduces to the estimator $t_3 = \overline{y}\exp\left(\frac{P-p}{P+p}\right)$ and for $\alpha = 0$, it reduces to

$$t_4 = \overline{y}\exp\left(\frac{p-P}{p+P}\right).$$

**Bias and MSE of $t_5$:**

Expressing (3.1) in terms of e's, we have

$$t_5 = \overline{Y}(1+e_y)\left[\alpha\exp\left\{\frac{P-P(1+e_\phi)}{P+P(1+e_\phi)}\right\} + (1-\alpha)\exp\left\{\frac{P(1+e_\phi)-P}{P(1+e_\phi)+P}\right\}\right]$$

$$= \overline{Y}(1+e_y)\left[\alpha\exp\left\{\frac{-e_\phi}{2}\right\} + (1-\alpha)\exp\left\{\frac{e_\phi}{2}\right\}\right] \quad (3.2)$$

Expanding the right hand side of (3.2) and retaining terms up to second powers of e's, we have

$$t_5 = \overline{Y}\left[1 + e_y + \frac{e_\phi}{2} - \alpha e_\phi + \frac{e_\phi^2}{8} + e_y e_\phi - \alpha e_y e_\phi\right] \quad (3.3)$$

Taking expectations of both sides of (3.3) and then subtracting $\overline{Y}$ from both sides, we get the bias of the estimator $t_5$ up to the first order of approximation, as

$$B(t_5) = f_1\overline{Y}\left[\frac{C_p^2}{8} + \rho_{pb}C_yC_p\left(\frac{1}{2}-\alpha\right)\right] \quad (3.4)$$

From (3.3), we have

$$(t_5 - \overline{Y}) \cong \overline{Y}\left[e_y + e_\phi\left(\frac{1}{2}-\alpha\right)\right] \quad (3.5)$$

Squaring both sides of (3.5) and then taking expectations we get MSE of the estimator $t_5$, up to the first order of approximation as

$$MSE(t_5) = f_1\overline{Y}^2\left[C_y^2 + C_p^2\left(\frac{1}{4}+\alpha^2-\alpha\right) + 2\rho_{pb}C_yC_p\left(\frac{1}{2}-\alpha\right)\right] \quad (3.6)$$

Minimization of (3.6) with respect to $\alpha$ yields optimum value of as

$$\alpha = \frac{2K_p+1}{2} = \alpha_0 \text{ (Say)} \quad (3.7)$$

Substitution of (3.7) in (3.1) yields the optimum estimator for $t_5$ as $(t_5)_{opt}$ (say) with minimum MSE as

$$\min.MSE(t_5) = f_1 \overline{Y}^2 C_y^2 (1 - \rho_{pb}^2) = M(t_5)_{opt} \qquad (3.8)$$

which is same as that of traditional linear regression estimator.

## 4. Efficiency comparisons

In this section, the conditions for which the proposed estimator $t_5$ is better than $\overline{y}$, $t_1$, $t_2$, $t_3$, and $t_4$ have been obtained. The variance of $\overline{y}$ is given by

$$\text{var}(\overline{y}) = f_1 \overline{Y}^2 C_y^2 \qquad (4.1)$$

To compare the efficiency of the proposed estimator $t_5$ with the existing estimator, from (4.1) and (1.3), (1.4), (2.5), (2.8) and (3.8), we have

$$\text{var}(\overline{y}) - M(t_5)_0 = \rho_{pb}^2 \geq 0. \qquad (4.2)$$

$$MSE(t_1) - M(t_5)_0 = (C_p - \rho_{pb} C_y)^2 \geq 0. \qquad (4.3)$$

$$MSE(t_2) - M(t_5)_0 = (C_p + \rho_{pb} C_y)^2 \geq 0. \qquad (4.4)$$

$$MSE(t_3) - M(t_5)_0 = \left(\frac{C_p^2}{2} - \rho_{pb} C_y\right)^2 \geq 0. \qquad (4.5)$$

$$MSE(t_4) - M(t_5)_0 = \left(\frac{C_p^2}{2} + \rho_{pb} C_y\right)^2 \geq 0. \qquad (4.6)$$

Using (4.2)-(4.6), we conclude that the proposed estimator $t_5$ outperforms $\overline{y}$, $t_1$, $t_2$, $t_3$, and $t_4$.

## 5. Empirical study

We now compare the performance of various estimators considered here using the following data sets:

**Population 1. [Source: Sukhatme and Sukhatme (1970), p. 256]**

y = number of villages in the circles and

ϕ = A circle consisting more than five villages.

N = 89, $\overline{Y}$ = 3.360, P = 0.1236, $\rho_{pb}$ = 0.766, $C_y$ = 0.60400, $C_p$ = 2.19012.

**Population 2. [Source: Mukhopadhyaya, (2000), p. 44]**

Y = Household size and

ϕ = A household that availed an agricultural loan from a bank.

N = 25, $\overline{Y}$ = 9.44, P = 0.400, $\rho_{pb}$ = -0.387, $C_y$ = 0.17028, $C_p$ = 1.27478.

The percent relative efficiency (PRE's) of the estimators $\overline{y}$, $t_1$-$t_4$ and $(t_5)_{opt}$ with respect to unusual unbiased estimator $\overline{y}$ have been computed and compiled in table 5.1.

**Table 5.1: PRE of various estimators with respect to $\overline{y}$.**

| Estimator | PRE's (., $\overline{y}$) | |
|---|---|---|
| | Population | |
| | I | II |
| $\overline{y}$ | 100 | 100 |
| $t_1$ | 11.63 | 1.59 |
| $t_2$ | 5.07 | 1.94 |
| $t_3$ | 66.24 | 5.57 |
| $t_4$ | 14.15 | 8.24 |
| $(t_5)_0$ | 241.98 | 117.61 |

Table 5.1 shows that the proposed estimator $t_5$ under optimum condition performs better than the usual sample mean $\bar{y}$, Naik and Gupta (1996) estimators ($t_1$ and $t_2$) and the ratio and product type exponential estimators ($t_3$ and $t_4$).

## 6. Double sampling

In some practical situations when P is not known a priori, the technique of two-phase sampling is used. Let p' denote the proportion of units possessing attribute $\phi$ in the first phase sample of size n'; p denote the proportion of units possessing attribute $\phi$ in the second phase sample of size n < n' and $\bar{y}$ denote the mean of the study variable y in the second phase sample.

When P is not known, two-phase ratio and product type exponential estimator are given by

$$t_6 = \bar{y}\exp\left(\frac{p'-p}{p'+p}\right) \tag{6.1}$$

$$t_7 = \bar{y}\exp\left(\frac{p-p'}{p+p'}\right) \tag{6.2}$$

To obtain the bias and MSE of $t_6$ and $t_7$, we write

$$\bar{y} = \bar{Y}(1+e_y), \quad p = P(1+e_\phi), \quad p' = P(1+e'_\phi)$$

such that

$$E(e_y) = E(e_\phi) = E(e'_\phi) = 0.$$

and

$$E(e_y^2) = f_1 C_y^2, \quad E(e_\phi^2) = f_1 C_p^2, \quad E(e'_\phi)^2 = f_2 C_p^2, \quad E(e_\phi e'_\phi) = f_2 \rho_{pb} C_y C_p.$$

where $f_2 = \dfrac{1}{n'} - \dfrac{1}{N}$.

Expressing (6.1) in terms of e's, we have

$$t_6 = \overline{Y}(1+e_y)\exp\left[\dfrac{P(1+e'_\phi) - P(1+e_\phi)}{P(1+e'_\phi) + P(1+e_\phi)}\right]$$

$$= \overline{Y}(1+e_y)\exp\left[\dfrac{e'_\phi - e_\phi}{2}\right] \qquad (6.3)$$

Expanding the right hand side of (6.3) and retaining terms up to second powers of e's, we have

$$t_6 = \overline{Y}\left[1 + e_y + \dfrac{e'_\phi}{2} - \dfrac{e_\phi}{2} + \dfrac{e'^2_\phi}{8} + \dfrac{e^2_\phi}{8} - \dfrac{e_\phi e'_\phi}{4} + \dfrac{e_y e'_\phi}{2} - \dfrac{e_y e_\phi}{2}\right] \qquad (6.4)$$

Taking expectations of both sides of (6.4) and then subtracting $\overline{Y}$ from both sides, we get the bias of the estimator $t_6$ up to the first order of approximation, as

$$B(t_6) = f_3 \overline{Y} \dfrac{C_p^2}{4}(1 - 2K_p) \qquad (6.5)$$

From (6.4), we have

$$(t_6 - \overline{Y}) \cong \overline{Y}\left[e_y + \dfrac{(e'_\phi - e_\phi)}{2}\right] \qquad (6.6)$$

Squaring both sides of (6.6) and then taking expectations we get MSE of the estimator $t_6$, up to the first order of approximation as

$$\text{MSE}(t_6) = \overline{Y}^2\left[f_1 C_y^2 + f_3 \dfrac{C_p^2}{4}(1 - 4K_p)\right] \qquad (6.7)$$

To obtain the bias and MSE of $t_7$ to the first degree of approximation, we express (6.2) in terms of e's as

$$t_7 = \overline{Y}(1+e_y)\exp\left[\frac{P(1+e_\phi)-P(1+e'_\phi)}{P(1+e_\phi)+P(1+e'_\phi)}\right]$$

$$= \overline{Y}(1+e_y)\exp\left[\frac{e_\phi - e'_\phi}{2}\right] \tag{6.8}$$

Expanding the right hand side of (6.8) and retaining terms up to second powers of e's, we have

$$t_7 = \overline{Y}\left[1+e_y+\frac{e_\phi}{2}-\frac{e'_\phi}{2}+\frac{e'^2_\phi}{8}+\frac{e^2_\phi}{8}-\frac{e_\phi e'_\phi}{4}+\frac{e_y e_\phi}{2}-\frac{e_y e'_\phi}{2}\right] \tag{6.9}$$

Taking expectations of both sides of (6.9) and then subtracting $\overline{Y}$ from both sides, we get the bias of the estimator $t_7$ up to the first order of approximation, as

$$B(t_7) = f_3 \overline{Y}\frac{C^2_p}{4}(1+2K_p) \tag{6.10}$$

From (6.9), we have

$$(t_7 - \overline{Y}) \cong \overline{Y}\left[e_y + \frac{(e_\phi - e'_\phi)}{2}\right] \tag{6.11}$$

Squaring both sides of (6.11) and then taking expectations we get MSE of the estimator $t_7$, up to the first order of approximation as

$$\text{MSE}(t_7) = \overline{Y}^2\left[f_1 C^2_y + f_3 \frac{C^2_p}{4}(1+4K_p)\right] \tag{6.12}$$

## 7. Proposed class of estimators in double sampling

We propose the following class of estimators in double sampling

$$t_8 = \overline{y}\left[\alpha_1 \exp\left(\frac{p'-p}{p'+p}\right) + (1-\alpha_1)\exp\left(\frac{p-p'}{p+p'}\right)\right] \tag{7.1}$$

where $\alpha_1$ is a real constant to be determined such that the MSE of $t_8$ is minimum.

For $\alpha_1 = 1$, $t_8$ reduces to the estimator $t_6 = \bar{y} \exp\left(\dfrac{p'-p}{p'+p}\right)$ and for $\alpha_1 = 0$, it reduces to

$$t_7 = \bar{y} \exp\left(\dfrac{p-p'}{p+p'}\right).$$

**Bias and MSE of $t_8$:**

Expressing (7.1) in terms of e's, we have

$$t_8 = \bar{Y}(1+e_y)\left[\alpha_1 \exp\left\{\dfrac{P(1+e'_\phi)-P(1+e\phi)}{P(1+e'_\phi)+P(1+e\phi)}\right\} + (1-\alpha_1)\exp\left\{\dfrac{P(1+e_\phi)-P(1+e'_\phi)}{P(1+e_\phi)+P(1+e'_\phi)}\right\}\right]$$

$$= \bar{Y}(1+e_y)\left[\alpha_1 \exp\left\{\dfrac{e'_\phi - e_\phi}{2}\right\} + (1-\alpha_1)\exp\left\{\dfrac{e_\phi - e'_\phi}{2}\right\}\right] \qquad (7.2)$$

Expanding the right hand side of (7.2) and retaining terms up to second powers of e's, we have

$$t_8 = \bar{Y}[1 + e_y + \dfrac{e_\phi}{2} - \dfrac{e'_\phi}{2} - \alpha_1 e_\phi + \alpha_1 e'_\phi + \dfrac{e_\phi^2}{8} + \dfrac{{e'_\phi}^2}{8} + \dfrac{e_y e_\phi}{2} - \dfrac{e_y e'_\phi}{2}$$

$$- \dfrac{e_\phi e'_\phi}{4} + \alpha_1 e_y e'_\phi - \alpha_1 e_y e_\phi] \qquad (7.3)$$

Taking expectations of both sides of (7.3) and then subtracting $\bar{Y}$ from both sides, we get the bias of the estimator $t_8$ up to the first order of approximation, as

$$B(t_8) = f_3 \bar{Y} \dfrac{C_p^2}{8}\left[1 - 8K_p\left(\alpha_1 - \dfrac{1}{2}\right)\right] \qquad (7.4)$$

From (7.3), we have

$$(t_8 - \bar{Y}) \cong \bar{Y}\left[e_y - \left(\alpha_1 - \dfrac{1}{2}\right)e_\phi + \left(\alpha_1 - \dfrac{1}{2}\right)e'_\phi\right] \qquad (7.5)$$

Squaring both sides of (7.5) and then taking expectations we get MSE of the estimator $t_8$, up to the first order of approximation as

$$\text{MSE}(t_8) = \overline{Y}^2\left[f_1 C_y^2 + f_3 C_p^2\left(\alpha_1 - \frac{1}{2}\right)\left\{\left(\alpha_1 - \frac{1}{2}\right) - 2K_p\right\}\right] \quad (7.6)$$

Minimization of (7.6) with respect to $\alpha_1$ yields optimum value of as

$$\alpha_1 = \frac{2K_p + 1}{2} = \alpha_{10}\,(\text{Say}) \quad (7.7)$$

Substitution of (7.7) in (7.1) yields the optimum estimator for $t_8$ as $(t_8)_{opt}$ (say) with minimum MSE as

$$\min.\text{MSE}(t_8) = \overline{Y}^2 C_y^2\left(f_1 - f_3 \rho_{pb}^2\right) = M(t_8)_o,\,(\text{say}) \quad (7.8)$$

which is same as that of traditional linear regression estimator.

## 8. Efficiency comparisons

The MSE of usual two-phase ratio and product estimator is given by

$$\text{MSE}(t_9) = \overline{Y}^2\left[f_1 C_y^2 + f_3 C_p^2(1 - 2K_p)\right] \quad (8.1)$$

$$\text{MSE}(t_{10}) = \overline{Y}^2\left[f_1 C_y^2 + f_3 C_p^2(1 + 2K_p)\right] \quad (8.2)$$

From (4.1), (6.7), (6.12), (8.1), (8.2) and (7.8) we have

$$\text{var}(\overline{y}) - M(t_8)_0 = f_3 \rho_{pb}^2 \geq 0. \quad (8.3)$$

$$\text{MSE}(t_6) - M(t_8)_0 = f_3\left(\frac{C_p}{2} - \rho_{pb} C_y\right)^2 \geq 0. \quad (8.4)$$

$$\text{MSE}(t_7) - M(t_8)_0 = f_3\left(\frac{C_p}{2} + \rho_{pb} C_y\right)^2 \geq 0. \quad (8.5)$$

$$\text{MSE}(t_9) - M(t_8)_0 = f_3\left(C_p - \rho_{pb} C_y\right)^2 \geq 0. \quad (8.6)$$

$$\mathrm{MSE}(t_{10}) - M(t_8)_0 = f_3\left(C_p + \rho_{pb}C_y\right)^2 \geq 0. \tag{8.7}$$

From (8.3)-(8.7), we conclude that our proposed estimator $t_8$ is better than $\bar{y}$, $t_6$, $t_7$, $t_9$, and $t_{10}$.

## 9. Empirical study

The various results obtained in the previous section are now examined with the help of following data:

**Population 1. [Source: Sukhatme and Sukhatme( 1970), p. 256]**

$N = 89$, $n' = 45$, $n = 23$, $\bar{y} = 1322$, $p = 0.1304$, $p' = 0.1333$, $\rho_{pb} = 0.408$, $C_y = 0.69144$, $C_p = 2.7005$.

**Population 2. [Source: Mukhopadhyaya( 2000), p. 44]**

$N = 25$, $n' = 13$, $n = 7$, $\bar{y} = 7.143$, $p = 0.294$, $p' = 0.308$, $\rho_{pb} = -0.314$, $C_y = 0.36442$, $Cp = 1.34701$.

**Table 9.1: PRE of various estimators (double sampling) with respect to $\bar{y}$.**

| Estimator | PRE's $(., \bar{y})$ | |
|---|---|---|
| | Population | |
| | I | II |
| $\bar{y}$ | 100 | 100 |
| $t_6$ | 40.59 | 25.42 |
| $t_7$ | 21.90 | 40.89 |
| $t_9$ | 11.16 | 8.89 |
| $t_{10}$ | 7.60 | 12.09 |
| $(t_8)_0$ | 112.32 | 106.74 |

Table 9.1 shows that the proposed estimator $t_8$ under optimum condition performs better than the usual sample mean $\bar{y}$, $t_6$, $t_7$, $t_9$, and $t_{10}$.

**References**

Bahl, S. and Tuteja, R.K. (1991): Ratio and Product type exponential estimator, Information and Optimization sciences, Vol.XII, I, 159-163.

Jhajj, H. S., Sharma, M. K. and Grover, L. K. (2006): A family of estimators of population mean using information on auxiliary attribute. Pak. J. Statist., 22 (1), 43-50.

Mukhopadhyaya, P.(2000): Theory and methods of survey sampling. Prentice Hall of India, New Delhi, India.

Naik,V.D. and Gupta, P.C. (1996): A note on estimation of mean with known population proportion of an auxiliary character. Jour. Ind. Soc. Agr. Stat., 48(2),151-158.

Shabbir,J. and Gupta, S.(2007) : On estimating the finite population mean with known population proportion of an auxiliary variable. Pak. J. Statist.,23(1),1-9.

Singh, H.P. and Espejo, M.R. (2003): On linear regression and ratio-product estimation of a finite population mean. The statistician, 52, 1, 59-67.

Sukhatme, P.V. and Sukhatme, B.V. (1970): Sampling theory of surveys with applications. Iowa State University Press, Ames, U.S.A.

# Improvement in Estimating the Population Mean Using Exponential Estimator in Simple Random Sampling

Rajesh Singh, Pankaj Chauhan, Nirmala Sawan

School of Statistics, DAVV, Indore (M.P.), India

(rsinghstat@yahoo.com)

Florentin Smarandache

Department of Mathematics, University of New Mexico, Gallup, USA

(fsmarandache@yahoo.com)

**Abstract**

This study proposes some exponential ratio-type estimators for estimating the population mean of the variable under study, using known values of certain population parameter(s). Under simple random sampling without replacement (SRSWOR) scheme, mean square error (MSE) equations of all proposed estimators are obtained and compared with each other. The theoretical results are supported by a numerical illustration.

**Keywords**: Exponential estimator, auxiliary variable, simple random sampling, efficiency.

## 1. Introduction

Consider a finite population $U = U_1, U_2, ....., U_N$ of N unites. Let y and x stand for the variable under study and auxiliary variable respectively. Let $(y_i, x_i)$, $i = 1, 2, ......, n$ denote the values of the units included in a sample $s_n$ of size n drawn by simple random sampling without replacement (SRSWOR). The auxiliary information has been used in improving the precision of the estimate of a parameter (see Cochran (1977), Sukhatme and Sukhatme (1970) and the reference cited there in). Out of many methods, ratio and product methods of estimation are good illustrations in this context.

In order to have a survey estimate of the population mean $\bar{Y}$ of the study character y, assuming the knowledge of the population mean $\bar{X}$ of the auxiliary character x, the well-known ratio estimator is –

$$t_r = \bar{y}\left(\frac{\bar{X}}{\bar{x}}\right) \tag{1.1}$$

Bahl and Tuteja (1991) suggested an exponential ratio type estimator as –

$$t_1 = \bar{y}\exp\left[\frac{\bar{X}-\bar{x}}{\bar{X}+\bar{x}}\right] \tag{1.2}$$

Several authors have used prior value of certain population parameter(s) to find more precise estimates. Sisodiya and Dwivedi (1981), Sen (1978) and Upadhyaya and Singh (1984) used the known coefficient of variation (CV) of the auxiliary character for estimating population mean of a study character in ratio method of estimation. The use of prior value of coefficient of kurtosis in estimating the population variance of study character y was first made by Singh et.al.(1973). Later used by Singh and Kakaran (1993) in the estimation of population mean of study character. Singh and Tailor (2003) proposed a modified ratio estimator by using the known value of correlation coefficient. Kadilar and Cingi (2006(a)) and Khoshnevisan et.al.(2007) have suggested modified ratio estimators by using different pairs of known value of population parameter(s).

In this paper, under SRSWOR, we have suggested improved exponential ratio-type estimators for estimating population mean using some known value of population parameter(s).

## 2. The suggested estimator

Following Kadilar and Cingi (2006(a)) and Khoshnevisan (2007), we define modified exponential estimator for estimating $\bar{Y}$ as –

$$t = \bar{y}\exp\left[\frac{(a\bar{X}+b)-(a\bar{x}+b)}{(a\bar{X}+b)+(a\bar{x}+b)}\right] \tag{2.1}$$

where $a(\neq 0)$, b are either real numbers or the functions of the known parameters of the auxiliary variable x such as coefficient of variation $(C_x)$, coefficient of kurtosis $(\beta_2(x))$ and correlation coefficient $(\rho)$.

To, obtain the bias and MSE of t, we write

$$\bar{y} = \bar{Y}(1+e_0), \quad \bar{x} = \bar{X}(1+e_1)$$

such that

$$E(e_0) = E(e_1) = 0,$$

and $E(e_0^2) = f_1 C_y^2$, $E(e_1^2) = f_1 C_x^2$, $E(e_0 e_1) = f_1 \rho C_y C_x$,

where

$$f_1 = \frac{N-n}{nN}, \quad C_y^2 = \frac{S_y^2}{\bar{Y}^2}, \quad C_x^2 = \frac{S_x^2}{\bar{X}^2}.$$

Expressing t in terms of e's, we have

$$t = \bar{Y}(1+e_0)\exp\left[\frac{a\bar{X} - a\bar{X}(1+e_1)}{a\bar{X} + 2b + a\bar{X}(1+e_1)}\right]$$

$$t = \bar{Y}(1+e_0)\exp\left[-\theta e_1(1+\theta e_1)^{-1}\right] \tag{2.2}$$

where $\theta = \dfrac{a\bar{X}}{2(a\bar{X}+b)}$.

Expanding the right hand side of (2.2) and retaining terms up to second power of e's, we have

$$t = \bar{Y}(1 + e_0 - \theta e_1 + \theta^2 e_1^2 + \theta e_0 e_1) \tag{2.3}$$

Taking expectations of both sides of (2.3) and then subtracting $\bar{Y}$ from both sides, we get the bias of the estimator t, up to the first order of approximation, as

$$B(t) = f_1 \bar{Y}(\theta^2 C_y^2 + \theta \rho C_y C_x) \tag{2.4}$$

From (2.3), we have

$$(t - \bar{Y}) \cong \bar{Y}(e_0 - \theta e_1) \tag{2.5}$$

Squaring both sides of (2.5) and then taking expectation, we get MSE of the estimator t, up to the first order of approximation, as

$$MSE(t) = f_1 \bar{Y}^2(C_y^2 + \theta^2 C_x^2 - 2\theta \rho C_y C_c) \tag{2.6}$$

### 3. Some members of the suggested estimator 't'

The following scheme presents some estimators of the population mean which can be obtained by suitable choice of constants a and b.

| Estimator | Values of | |
|---|---|---|
| | a | b |
| $t_0 = \bar{y}$<br>The mean per unit estimator | 0 | 0 |
| $t_1 = \bar{y}\exp\left[\dfrac{\bar{X}-\bar{x}}{\bar{X}+\bar{x}}\right]$<br>Bahl and Tuteja (1991) estimator | 1 | 1 |
| $t_2 = \bar{y}\exp\left[\dfrac{\bar{X}-\bar{x}}{\bar{X}+\bar{x}+2\beta_2(x)}\right]$ | 1 | $\beta_2(x)$ |
| $t_3 = \bar{y}\exp\left[\dfrac{\bar{X}-\bar{x}}{\bar{X}+\bar{x}+2C_c}\right]$ | 1 | $C_x$ |
| $t_4 = \bar{y}\exp\left[\dfrac{\bar{X}-\bar{x}}{\bar{X}+\bar{x}+2\rho}\right]$ | 1 | $\rho$ |
| $t_5 = \bar{y}\exp\left[\dfrac{\beta_2(x)(\bar{X}-\bar{x})}{\beta_2(x)(\bar{X}+\bar{x})+2C_x}\right]$ | $\beta_2(x)$ | $C_x$ |
| $t_6 = \bar{y}\exp\left[\dfrac{C_x(\bar{X}-\bar{x})}{C_x(\bar{X}+\bar{x})+2\beta_2(x)}\right]$ | $C_x$ | $\beta_2(x)$ |
| $t_7 = \bar{y}\exp\left[\dfrac{C_x(\bar{X}-\bar{x})}{C_x(\bar{X}+\bar{x})+2\rho}\right]$ | $C_x$ | $\rho$ |
| $t_8 = \bar{y}\exp\left[\dfrac{\rho(\bar{X}-\bar{x})}{\rho(\bar{X}+\bar{x})+2C_x}\right]$ | $\rho$ | $C_x$ |
| $t_9 = \bar{y}\exp\left[\dfrac{\beta_2(x)(\bar{X}-\bar{x})}{\beta_2(x)(\bar{X}+\bar{x})+2\rho}\right]$ | $\beta_2(x)$ | $\rho$ |
| $t_{10} = \bar{y}\exp\left[\dfrac{\rho(\bar{X}-\bar{x})}{\rho(\bar{X}+\bar{x})+2\beta_2(x)}\right]$ | $\rho$ | $\beta_2(x)$ |

In addition to above estimators a large number of estimators can also be generated from the proposed estimator t at (2.1) just by putting different values of a and b.

It is observed that the expressions of the first order of approximation of bias and MSE of the given member of the family can be obtained by mere substituting the values of a and b in (2.4) and (2.6) respectively.

## 4. Modified estimators

Following Kadilar and Cingi (2006(b)), we propose modified estimator combining estimator $t_1$ and $t_i$ $(i=2,3,....,10)$ as follows

$$t_i^* = \alpha t_1 + (1-\alpha)t_i, \qquad (i=2,3,....,10) \tag{4.1}$$

where $\alpha$ is a real constant to be determined such that the MSE of $t_i^*$ is minimum and $t_i$ $(i=2,3,....,10)$ are estimators listed in section 3.

Following the procedure of section (2), we get the MSE of $t_i^*$ to the first order of approximation as –

$$\text{MSE}(t_i^*) = f_1 \overline{Y}^2 \left[ C_y^2 + C_x^2 \left( \frac{\alpha}{2} + \theta_i - \alpha \theta_i \right)^2 - 2\rho C_y C_x \left( \frac{\alpha}{2} + \theta_i - \alpha \theta_i \right) \right] \tag{4.2}$$

where

$$\theta_2 = \frac{\overline{X}}{2(\overline{X}+\beta_2(x))}, \qquad \theta_3 = \frac{\overline{X}}{2(\overline{X}+C_x)},$$

$$\theta_4 = \frac{\overline{X}}{2(\overline{X}+\rho)}, \qquad \theta_5 = \frac{\beta_2(x)\overline{X}}{2(\beta_2(x)\overline{X}+C_x)},$$

$$\theta_6 = \frac{C_x \overline{X}}{2(C_x \overline{X}+\beta_2(x))}, \qquad \theta_7 = \frac{C_x \overline{X}}{2(C_x \overline{X}+\rho)},$$

$$\theta_8 = \frac{\rho \overline{X}}{2(\rho \overline{X}+C_x)}, \qquad \theta_9 = \frac{\beta_2(x)\overline{X}}{2(\beta_2(x)+\rho)},$$

$$\theta_{10} = \frac{\rho \overline{X}}{2(\rho+\beta_2(x))}.$$

Minimization of (4.2) with respect to $\alpha$ yields its optimum value as

$$\alpha = \frac{2(K-\theta)}{(1-2\theta)} = \alpha_{opt} \text{ (say)} \tag{4.3}$$

where $K = \rho \dfrac{C_y}{C_x}$.

Substitution of (4.3) in (4.10) gives optimum estimator $t_o^*$ (say), with minimum MSE as

$$\min \mathrm{MSE}(t_i^*) = f_1 \overline{Y}^2 C_y^2 (1-\rho^2) = \mathrm{MSE}(t^*)_o \tag{4.4}$$

The $\min \mathrm{MSE}(t_i^*)$ at (4.4) is same as that of the approximate variance of the usual linear regression estimator.

## 5. Efficiency comparison

It is well known that under SRSWOR the variance of the sample mean is

$$\mathrm{Var}(\overline{y}) = f_1 \overline{Y}^2 C_y^2 \tag{5.1}$$

we first compare the MSE of the proposed estimators, given in (2.6) with the variance of the sample mean, we have the following condition:

$$K \leq \frac{\theta_i}{2}, \qquad i = 2,3,\ldots,10 \tag{5.2}$$

When this condition is satisfied, proposed estimators are more efficient than the sample mean.

Next we compare the MSE of proposed estimators $t_i^*$ ($i=2,3,\ldots,10$) in (4.4) with the MSE of estimators listed in section 3. We obtain the following condition

$$(\theta C_x - \rho C_y)^2 \geq 0, \qquad i = 2,3,\ldots,10. \tag{5.3}$$

We can infer that all proposed estimators $t_i^*$, ($i=2,3,\ldots,10$) are more efficient than estimators proposed in section 3 in all conditions, because the condition given in (5.1) is always satisfied.

## 6. Numerical illustration

To illustrate the performance of various estimators of $\overline{Y}$, we consider the data given in Murthy (1967 pg-226). The variates are:

y : Output,    x: number of workers

$\overline{X} = 283.875$, $\overline{Y} = 5182.638$, $C_y = 0.3520$, $C_x = 0.9430$, $\rho = 0.9136$, $\beta_2(x) = 3.65$.

We have computed the percent relative efficiency (PRE) of different estimators of $\overline{Y}$ with respect to usual estimator $\overline{y}$ and complied in table 6.1:

**Table 6.1: PRE of different estimators of $\overline{Y}$ with respect to $\overline{y}$**

| Estimator | PRE |
|---|---|
| $\overline{y}$ | 100 |
| $t_1$ | 366.96 |
| $t_2$ | 385.72 |
| $t_3$ | 368.27 |
| $t_4$ | 371.74 |
| $t_5$ | 386.87 |
| $t_6$ | 368.27 |
| $t_7$ | 372.03 |
| $t_8$ | 372.05 |
| $t_9$ | 368.27 |
| $t_{10}$ | 386.91 |
| $t_o^*$ | 877.54 |

## 7. Conclusion

We have developed some exponential ratio type estimators using some known value of the population parameter(s), listed in section 3. We have also suggested modified estimators $t_i^*$ ($i = 2,3,\ldots,10$). From table 6.1 we conclude that the proposed estimators are better than Bahl and Tuteja (1991) estimator $t_1$. Also, the modified estimator $t_i^*$ ($i = 2,3,\ldots,10$) under optimum condition performs better than the estimators proposed and listed in section 3 and than the Bahl and Tuteja (1991) estimator $t_1$. The choice of the estimator mainly depends upon the availability of information about known values of the parameter(s) ($C_x$, $\rho$, $\beta_2(x)$, etc.).

**References**

Bahl, S. and Tuteja, R.K. (1991): Ratio and Product type exponential estimator, Information and Optimization sciences, Vol.XII, I, 159-163.

Cochran, W.G.(1977): Sampling techniques. Third U.S. edition. Wiley eastern limited, 325.

Kadilar, C. and Cingi, H. (2006(a)): A new ratio estimator using correlation coefficient. Inter Stat, 1-11.

Kadilar, C. and Cingi, H. (2006(b)): Improvement in estimating the population mean in simple random sampling. Applied Mathematics Letters, 19,75-79.

Khoshnevisan, M. Singh, R., Chauhan, P., Sawan, N. and Smarandache, F. (2007): A general family of estimators for estimating population mean using known value of some population parameter(s). Far east journal of theoretical statistics, 22(2), 181-191.

Murthy, M.N., (1967): Sampling Theory and Methods, Statistical Publishing Society, Calcutta.

Sen, A.R., (1978) : Estimation of the population mean when the coefficient of variation is known. Comm. Stat.-Theory Methods, A7, 657-672.

Singh, H.P. and Kakran, M.S. (1993): A modified ratio estimator using coefficient of variation of auxiliary character. Unpublished.

Singh, H.P. and Tailor, R. (2003): Use of known correlation coefficient in estimating the finite population mean. Statistics in Transition, 6,4,555-560.

Singh, J. Pandey, B.N. and Hirano, K. (1973): On the utilization of a known coefficient of kurtosis in the estimation procedure of variance. Ann. Inst. Stat. Math., 25, 51-55.

Sisodiya, B.V.S. and Dwivedi, V.K. (1981): A modified ratio estimator using coefficient of variation of auxiliary variable. Jour. Ind. Soc. Agri. Stat., 33, 13-18.

Sukhatme, P.V. and Sukhatme, B.V., (1970): Sampling theory of surveys with applications. Iowa State University Press, Ames, U.S.A.

Upadhyaya, L.N. and Singh, H.P.,(1984): On the estimation of the population mean with known coefficient of variation. Biometrical Journal, 26, 915-922.

# Almost Unbiased Exponential Estimator for the Finite Population Mean

Rajesh Singh, Pankaj Chauhan, Nirmala Sawan

School of Statistics, DAVV, Indore (M.P.), India

(rsinghstat@yahoo.com)

Florentin Smarandache

Chair of Department of Mathematics, University of New Mexico, Gallup, USA

(smarand@unm.edu)

**Abstract**

In this paper we have proposed an almost unbiased ratio and product type exponential estimator for the finite population mean $\overline{Y}$. It has been shown that Bahl and Tuteja (1991) ratio and product type exponential estimators are particular members of the proposed estimator. Empirical study is carried to demonstrate the superiority of the proposed estimator.

**Keywords**: Auxiliary information, bias, mean-squared error, exponential estimator.

## 1. Introduction

It is well known that the use of auxiliary information in sample surveys results in substantial improvement in the precision of the estimators of the population mean. Ratio, product and difference methods of estimation are good examples in this context. Ratio method of estimation is quite effective when there is a high positive correlation between study and auxiliary variables. On other hand, if this correlation is negative (high), the product method of estimation can be employed effectively.

Consider a finite population with N units $(U_1, U_2, ...., U_N)$ for each of which the information is available on auxiliary variable x. Let a sample of size n be drawn with simple random sampling without replacement (SRSWOR) to estimate the population mean of character y under study. Let $(\bar{y}, \bar{x})$ be the sample mean estimator of $(\bar{Y}, \bar{X})$ the population means of y and x respectively.

In order to have a survey estimate of the population mean $\bar{Y}$ of the study character y, assuming the knowledge of the population mean $\bar{X}$ of the auxiliary character x, Bahl and Tuteja (1991) suggested ratio and product type exponential estimator

$$t_1 = \bar{y} \exp\left(\frac{\bar{X} - \bar{x}}{\bar{X} + \bar{x}}\right) \qquad (1.1)$$

$$t_2 = \bar{y} \exp\left(\frac{\bar{x} - \bar{X}}{\bar{x} + \bar{X}}\right) \qquad (1.2)$$

Up to the first order of approximation, the bias and mean-squared error (MSE) of $t_1$ and $t_2$ are respectively given by

$$B(t_1) = \left(\frac{N-n}{nN}\right) \bar{Y} \frac{C_x^2}{2}\left(\frac{1}{2} - K\right) \qquad (1.3)$$

$$MSE(t_1) = \left(\frac{N-n}{nN}\right) \bar{Y}^2 \left[C_y^2 + C_x^2\left(\frac{1}{4} - K\right)\right] \qquad (1.4)$$

$$B(t_2) = \left(\frac{N-n}{nN}\right) \bar{Y} \frac{C_x^2}{2}\left(\frac{1}{2} + K\right) \qquad (1.5)$$

$$\text{MSE}(t_2) = \left(\frac{N-n}{nN}\right)\overline{Y}^2\left[C_y^2 + C_x^2\left(\frac{1}{4} + K\right)\right] \tag{1.6}$$

where $S_y^2 = \frac{1}{(N-1)}\sum_{i=1}^{N}(y_i - \overline{Y})^2$, $S_x^2 = \frac{1}{(N-1)}\sum_{i=1}^{N}(x_i - \overline{X})^2$, $C_y = \frac{S_y}{\overline{Y}}$, $C_x = \frac{S_x}{\overline{X}}$,

$K = \rho\left(\frac{C_y}{C_x}\right)$, $\rho = \frac{S_{yx}}{(S_y S_x)}$, $S_{yx} = \frac{1}{(N-1)}\sum_{i=1}^{N}(y_i - \overline{Y})(x_i - \overline{X})$.

From (1.3) and (1.5), we see that the estimators $t_1$ and $t_2$ suggested by Bahl and Tuteja (1991) are biased estimator. In some applications bias is disadvantageous. Following Singh and Singh (1993) and Singh and Singh (2006) we have proposed almost unbiased estimators of $\overline{Y}$.

## 2. Almost unbiased estimator

Suppose $t_0 = \overline{y}$, $t_1 = \overline{y}\exp\left(\frac{\overline{X} - \overline{x}}{\overline{X} + \overline{x}}\right)$, $t_2 = \overline{y}\exp\left(\frac{\overline{x} - \overline{X}}{\overline{x} + \overline{X}}\right)$

such that $t_0, t_1, t_2 \in H$, where H denotes the set of all possible estimators for estimating the population mean $\overline{Y}$. By definition, the set H is a linear variety if

$$t_h = \sum_{i=0}^{2} h_i t_i \in H \tag{2.1}$$

for $\sum_{i=0}^{2} h_i = 1$, $h_i \in R$ \hfill (2.2)

where $h_i (i = 0,1,2)$ denotes the statistical constants and R denotes the set of real numbers.

To obtain the bias and MSE of $t_h$, we write

$\overline{y} = \overline{Y}(1 + e_0)$, $\overline{x} = \overline{X}(1 + e_1)$.

such that

$E(e_0) = E(e_1) = 0$.

$$E(e_0^2) = \left(\frac{N-n}{Nn}\right)C_y^2, \quad E(e_1^2) = \left(\frac{N-n}{Nn}\right)C_x^2, \quad E(e_0 e_1) = \left(\frac{N-n}{Nn}\right)\rho C_y C_x.$$

Expressing $t_h$ in terms of e's, we have

$$t_h = \overline{Y}(1+e_0)\left[h_0 + h_1 \exp\left(\frac{-e_1}{2+e_1}\right) + h_2 \exp\left(\frac{e_1}{2+e_1}\right)\right] \quad (2.3)$$

Expanding the right hand side of (2.3) and retaining terms up to second powers of e's, we have

$$t_h = \overline{Y}\left[1 + e_0 - \frac{e_1}{2}(h_1 - h_2) + h_1\frac{e_1^2}{8} + h_2\frac{e_1^2}{8} - h_1\frac{e_0 e_1}{2} + h_2\frac{e_0 e_1}{2}\right] \quad (2.4)$$

Taking expectations of both sides of (2.4) and then subtracting $\overline{Y}$ from both sides, we get the bias of the estimator $t_h$, up to the first order of approximation as

$$B(t_h) = \left(\frac{N-n}{Nn}\right)\overline{Y}\frac{C_x^2}{2}\left[\frac{1}{4}(h_1 + h_2) - K(h_1 - h_2)\right] \quad (2.5)$$

From (2.4), we have

$$(t_h - \overline{Y}) \cong \overline{Y}\left[e_0 - h\frac{e_1}{2}\right] \quad (2.6)$$

where $h = h_1 - h_2$. $\quad (2.7)$

Squaring both the sides of (2.7) and then taking expectations, we get MSE of the estimator $t_h$, up to the first order of approximation, as

$$MSE(t_h) = \left(\frac{N-n}{Nn}\right)\overline{Y}^2\left[C_y^2 + C_x^2 h\left(\frac{h}{4} - K\right)\right] \quad (2.8)$$

which is minimum when

$$h = 2K. \quad (2.9)$$

Putting this value of $h = 2K$ in (2.1) we have optimum value of estimator as $t_h$ (optimum).

Thus the minimum MSE of $t_h$ is given by

$$\min.\text{MSE}(t_h) = \left(\frac{N-n}{Nn}\right)\bar{Y}^2 C_y^2 (1-\rho^2) \qquad (2.10)$$

which is same as that of traditional linear regression estimator.

From (2.7) and (2.9), we have

$$h_1 - h_2 = h = 2K. \qquad (2.11)$$

From (2.2) and (2.11), we have only two equations in three unknowns. It is not possible to find the unique values for $h_i$'s, i=0,1,2. In order to get unique values of $h_i$'s, we shall impose the linear restriction

$$\sum_{i=0}^{2} h_i B(t_i) = 0. \qquad (2.12)$$

where $B(t_i)$ denotes the bias in the $i^{th}$ estimator.

Equations (2.2), (2.11) and (2.12) can be written in the matrix form as

$$\begin{bmatrix} 1 & 1 & 1 \\ 0 & 1 & -1 \\ 0 & B(t_1) & B(t_2) \end{bmatrix} \begin{bmatrix} h_0 \\ h_1 \\ h_2 \end{bmatrix} = \begin{bmatrix} 1 \\ 2K \\ 0 \end{bmatrix} \qquad (2.13)$$

Using (2.13), we get the unique values of $h_i$'s(i=0,1,2) as

$$\left.\begin{aligned} h_0 &= 1 - 4K^2 \\ h_1 &= K + 2K^2 \\ h_2 &= -K + 2K^2 \end{aligned}\right\} \qquad (2.14)$$

Use of these $h_i$'s (i=0,1,2) remove the bias up to terms of order $o(n^{-1})$ at (2.1).

### 3. Two phase sampling

When the population mean $\bar{X}$ of x is not known, it is often estimated from a preliminary large sample on which only the auxiliary characteristic is observed. The value of population mean $\overline{X}$ of the auxiliary character x is then replaced by this estimate. This technique is known as the double sampling or two-phase sampling.

The two-phase sampling happens to be a powerful and cost effective (economical) procedure for finding the reliable estimate in first phase sample for the unknown parameters of the auxiliary variable x and hence has eminent role to play in survey sampling, for instance, see; Hidiroglou and Sarndal (1998).

When $\overline{X}$ is unknown, it is sometimes estimated from a preliminary large sample of size $n'$ on which only the characteristic x is measured. Then a second phase sample of size $n(n < n')$ is drawn on which both y and x characteristics are measured. Let $\overline{x}' = \frac{1}{n'}\sum_{i=1}^{n'} x_i$ denote the sample mean of x based on first phase sample of size $n'$; $\overline{y} = \frac{1}{n}\sum_{i=1}^{n} y_i$ and $\overline{x} = \frac{1}{n}\sum_{i=1}^{n} x_i$ be the sample means of y and x respectively based on second phase of size n.

In double (or two-phase) sampling, we suggest the following modified exponential ratio and product estimators for $\overline{Y}$, respectively, as

$$t_{1d} = \overline{y}\exp\left(\frac{\overline{x}' - \overline{x}}{\overline{x}' + \overline{x}}\right) \qquad (3.1)$$

$$t_{2d} = \overline{y}\exp\left(\frac{\overline{x} - \overline{x}'}{\overline{x} + \overline{x}'}\right) \qquad (3.2)$$

To obtain the bias and MSE of $t_{1d}$ and $t_{2d}$, we write

$$\overline{y} = \overline{Y}(1 + e_0), \ \overline{x} = \overline{X}(1 + e_1), \ \overline{x}' = \overline{X}(1 + e_1')$$

such that

$$E(e_0) = E(e_1) = E(e_1') = 0$$

and

$$E(e_0^2) = f_1 C_y^2, \qquad E(e_1^2) = f_1 C_x^2, \qquad E(e_1'^2) = f_2 C_x^2,$$

$$E(e_0 e_1) = f_1 \rho C_y C_x,$$

$$E(e_0 e_1') = f_2 \rho C_y C_x,$$

$$E(e_1 e_1') = f_2 C_x^2.$$

where $f_1 = \left(\dfrac{1}{n} - \dfrac{1}{N}\right)$, $f_2 = \left(\dfrac{1}{n'} - \dfrac{1}{N}\right)$.

Following standard procedure we obtain

$$B(t_{1d}) = \overline{Y} f_3 \left[ \frac{C_x^2}{8} - \frac{1}{2}\rho C_y C_x \right] \tag{3.3}$$

$$B(t_{2d}) = \overline{Y} f_3 \left[ \frac{C_x^2}{8} + \frac{1}{2}\rho C_y C_x \right] \tag{3.4}$$

$$MSE(t_{1d}) = \overline{Y}^2 \left[ f_1 C_y^2 + f_3 \left( \frac{C_x^2}{4} - \rho C_x C_y \right) \right] \tag{3.5}$$

$$MSE(t_{2d}) = \overline{Y}^2 \left[ f_1 C_y^2 + f_3 \left( \frac{C_x^2}{4} + \rho C_x C_y \right) \right] \tag{3.6}$$

where $f_3 = \left(\dfrac{1}{n} - \dfrac{1}{n'}\right)$.

From (3.3) and (3.4) we observe that the proposed estimators $t_{1d}$ and $t_{2d}$ are biased, which is a drawback of an estimator is some applications.

## 4. Almost unbiased two-phase estimator

Suppose $t_0 = \bar{y}$, $t_{1d}$ and $t_{2d}$ as defined in (3.1) and (3.2) such that $t_0, t_{1d}, t_{2d} \in W$, where W denotes the set of all possible estimators for estimating the population mean $\bar{Y}$. By definition, the set W is a linear variety if

$$t_W = \sum_{i=0}^{2} w_i t_i \in W. \tag{4.1}$$

for $\sum_{i=1}^{2} w_i = 1$, $w_i \in R$. \hfill (4.2)

where $w_i (i = 0,1,2)$ denotes thee statistical constants and R denotes the set of real numbers.

To obtain the bias and MSE of $t_w$, using notations of section 3 and expressing $t_w$ in terms of e's, we have

$$t_w = \bar{Y}(1+e_0)\left[w_0 + w_1 \exp\left(\frac{e'_1 - e_1}{2}\right) + w_2 \exp\left(\frac{e_1 - e'_1}{2}\right)\right] \tag{4.3}$$

$$t_w = \bar{Y}[1 + e_0 - \frac{w}{2}(e_1 - e'_1) + \frac{w_1}{8}(e_1^2 + e'^2_1) + \frac{w_2}{8}(e_1^2 + e'^2_1) - \left(\frac{w_1}{4} + \frac{w_2}{4}\right)e_1 e'_1$$

$$+ \frac{w}{2}(e_0 e'_1 - e_0 e_1)] \tag{4.4}$$

where $w = w_1 - w_2$. \hfill (4.5)

Taking expectations of both sides of (4.4) and then subtracting $\bar{Y}$ from both sides, we get the bias of the estimator $t_w$, up to the first order f approximation as

$$\text{Bias}(t_w) = \bar{Y} f_3 \left[\left(\frac{w_1 + w_2}{8}\right) C_x^2 - \frac{w}{2} \rho C_y C_x\right] \tag{4.6}$$

From (4.4), we have

$$t_w \cong \overline{Y}\left[e_0 - \frac{w}{2}(e_1 - e_1')\right] \qquad (4.7)$$

Squaring both sides of (4.7) and then taking expectation, we get MSE of the estimator $t_w$, up to the first order of approximation, as

$$MSE(t_w) = \overline{Y}^2\left[f_1 C_y^2 + f_3 w C_x^2\left(\frac{w}{4}\right) - K\right] \qquad (4.8)$$

which is minimum when

$$w = 2K. \qquad (4.9)$$

Thus the minimum MSE of $t_w$ is given by –

$$\min. MSE(t_w) = \overline{Y}^2 C_y^2 \left[f_1 - f_3 \rho^2\right] \qquad (4.10)$$

which is same as that of two-phase linear regression estimator. From (4.5) and (4.9), we have

$$w_1 - w_2 = w = 2K \qquad (4.11)$$

From (4.2) and (4.11), we have only two equations in three unknowns. It is not possible to find the unique values for $w_i$'s $(i = 0,1,2)$. In order to get unique values of $h_i$'s, we shall impose the linear restriction

$$\sum_{i=0}^{2} w_i B(t_{id}) = 0 \qquad (4.12)$$

where $B(t_{id})$ denotes the bias in the $i^{th}$ estimator.

Equations (4.2), (4.11) and (4.12) can be written in the matrix form as

$$\begin{bmatrix} 1 & 1 & 1 \\ 0 & 1 & -1 \\ 0 & B(t_{1d}) & B(t_{2d}) \end{bmatrix} \begin{bmatrix} w_0 \\ w_1 \\ w_2 \end{bmatrix} = \begin{bmatrix} 1 \\ 2K \\ 0 \end{bmatrix} \qquad (4.13)$$

Solving (4.13), we get the unique values of $w_i$'s $(i = 0,1,2)$ as –

$$\left.\begin{array}{l}w_0 = 1 - 8K^2 \\ w_1 = K + 4K^2 \\ w_2 = -K + 4K^2\end{array}\right\} \quad (4.14)$$

Use of these $w_i$'s $(i = 0,1,2)$ removes the bias up to terms of order $o(n^{-1})$ at (4.1).

## 5. Empirical study

The data for the empirical study are taken from two natural population data sets considered by Cochran (1977) and Rao (1983).

**Population I**: Cochran (1977)

$C_y = 1.4177$, $C_x = 1.4045$, $\rho = 0.887$.

**Population II**: Rao (1983)

$C_y = 0.426$, $C_x = 0.128$, $\rho = -0.7036$.

In table (5.1), the values of scalar $h_i$'s ($i = 0,1,2$) are listed.

**Table (5.1): Values of $h_i$'s ($i = 0,1,2$)**

| Scalars | Population | |
|---|---|---|
| | I | II |
| $h_0$ | -2.2065 | -20.93 |
| $h_1$ | 2.4985 | 8.62 |
| $h_2$ | 0.7079 | 13.30 |

Using these values of $h_i$'s ($i = 0,1,2$) given in the table 5.1, one can reduce the bias to the order $o(n^{-1})$ in the estimator $t_h$ at (2.1).

In table 5.2, Percent relative efficiency (PRE) of $\bar{y}$, $t_1$, $t_2$ and $t_h$ (in optimum case) are computed with respect to $\bar{y}$.

**Table 5.2: PRE of different estimators of $\overline{Y}$ with respect to $\bar{y}$.**

| Estimators | PRE ($.,\bar{y}$) | |
|---|---|---|
| | **Population I** | **Population II** |
| $\bar{y}$ | 100 | 100 |
| $t_1$ | 272.75 | 32.55 |
| $t_2$ | 47.07 | 126.81 |
| $t_h$ (optimum) | 468.97 | 198.04 |

Table 5.2 clearly shows that the suggested estimator $t_h$ in its optimum condition is better than usual unbiased estimator $\bar{y}$, Bahl and Tuteja (1991) estimators $t_1$ and $t_2$. For the purpose of illustration for two-phase sampling, we consider following populations:

**Population III:** Murthy (1967)

    y : Output
    x : Number of workers
$C_y = 0.3542$, $C_x = 0.9484$, $\rho = 0.9150$, $N = 80$, $n' = 20$, $n = 8$.

**Population IV:** Steel and Torrie(1960)

$C_y = 0.4803$, $C_x = 0.7493$, $\rho = -0.4996$, $N = 30$, $n' = 12$, $n = 4$.

In table 5.3 the values of scalars $w_i\text{'s}(i = 0,1,2)$ are listed.

**Table 5.3: Values of** $w_i\text{'s}(i = 0,1,2)$

| Scalars | Population I | Population II |
|---|---|---|
| $w_0$ | 0.659 | 0.2415 |
| $w_1$ | 0.808 | 0.0713 |
| $w_2$ | 0.125 | 0.6871 |

Using these values of $w_i\text{'s}(i = 0,1,2)$ given in table 5.3 one can reduce the bias to the order $o(n^{-1})$ in the estimator $t_w$ at 5.3.

In table 5.4 percent relative efficiency (PRE) of $\bar{y}$, $t_{1d}$, $t_{2d}$ and $t_w$ (in optimum case) are computed with respect to $\bar{y}$.

**Table 5.4: PRE of different estimators of** $\bar{Y}$ **with respect to** $\bar{y}$.

| Estimators | PRE $(.,\bar{y})$ | |
|---|---|---|
| | Population I | Population II |
| $\bar{y}$ | 100 | 100 |
| $t_{1d}$ | 128.07 | 74.68 |
| $t_{2d}$ | 41.42 | 103.64 |
| $t_w$ | 138.71 | 106.11 |

**References**

Bahl, S. and Tuteja, R.K. (1991): Ratio and product type exponential estimator. Information and optimization sciences, 12 (1), 159-163.

Cochran, W. G. (1977): Sampling techniques. Third edition Wiley and Sons, New York.

Hidiroglou, M. A. and Sarndal, C.E. (1998): Use of auxiliary information for two-phase sampling. Survey Methodology, 24(1), and 11—20.

Murthy, M..N (1967): Sampling Theory and Methods. Statistical Publishing Society, Calcutta, India.

Rao, T. J. (1983): A new class of unbiased product estimators. Tech. Rep. No. 15183, Stat. Math. Indian Statistical Institute, Calcutta, India.

Singh, R. and Singh, J. (2006): Separation of bias in the estimators of population mean using auxiliary information. Jour. Rajasthan Acad. Phy. Science, 5, 3, 327-332.

Singh, S. and Singh, R. (1993): A new method: Almost separation of bias precipitates in sample surveys. . Jour. Ind. Stat. Assoc., 31, 99-105.

Steel, R. G. D. and Torrie, J. H. (1960): Principles and procedures of statistics, Mc Graw Hill, New York.

# Almost Unbiased Ratio and Product Type Estimator of Finite Population Variance Using the Knowledge of Kurtosis of an Auxiliary Variable in Sample Surveys

Rajesh Singh, Pankaj Chauhan, Nirmala Sawan

School of Statistics, DAVV, Indore (M.P.), India

(rsinghstat@yahoo.com)

Florentin Smarandache

Chair of Department of Mathematics, University of New Mexico, Gallup, USA

(smarand@unm.edu)

**Abstract**

It is well recognized that the use of auxiliary information in sample survey design results in efficient estimators of population parameters under some realistic conditions. Out of many ratio, product and regression methods of estimation are good examples in this context. Using the knowledge of kurtosis of an auxiliary variable Upadhyaya and Singh (1999) have suggested an estimator for population variance. In this paper, following the approach of Singh and Singh (1993), we have suggested almost unbiased ratio and product-type estimators for population variance.

## 1. Introduction

Let $U = (U_1, U_2, \ldots, U_N)$ denote a population of N units from which a simple random sample without replacement (SRSWOR) of size n is to be drawn. Further let y and x denote the study and the auxiliary variables respectively. The problem is to estimate the parameter

$$S_y^2 = \frac{N}{N-1} \sigma_y^2 \qquad (1.1)$$

with $\sigma_y^2 = \frac{1}{N} \sum_{i=1}^{N} (y_i - \bar{Y})^2$ of the study variate y when the parameter

$$S_x^2 = \frac{N}{N-1} \sigma_x^2 \qquad (1.2)$$

with $\sigma_x^2 = \frac{1}{N} \sum_{i=1}^{N} (x_i - \bar{X})^2$ of the auxiliary variate x is known,

where $\bar{Y} = \sum_{i=1}^{N} \frac{y_i}{N}$ and $\bar{X} = \sum_{i=1}^{N} \frac{x_i}{N}$; are the population means of y and x respectively.

The conventional unbiased estimator of $S_y^2$ is defined by

$$s_y^2 = \frac{\sum_{i=1}^{n} (y_i - \bar{Y})^2}{(n-1)} \qquad (1.3)$$

where $\bar{y} = \sum_{i=1}^{n} \frac{y_i}{n}$ is the sample mean of y.

Using information on $S_x^2$, Isaki (1983) proposed a ratio estimator for $S_y^2$ as

$$t_1 = s_y^2 \frac{S_x^2}{s_x^2} \qquad (1.4)$$

where $s_x^2 = \frac{1}{(n-1)} \sum_{i=1}^{n} (x_i - \bar{x})^2$ is unbiased estimator of $S_x^2$.

In many survey situations the values of the auxiliary variable x may be available for each unit in the population. Thus the value of the kurtosis $\beta_2(x)$ of the auxiliary variable x is known. Using information on both $S_x^2$ and $\beta_2(x)$ Upadhyay and Singh (1999) suggested a ratio type estimator for $S_y^2$ as

$$t_2 = s_y^2 \left[ \frac{S_x^2 + \beta_2(x)}{s_x^2 + \beta_2(x)} \right] \tag{1.5}$$

For simplicity suppose that the population size N is large enough relative to the sample size n and assume that the finite population correction (fpc) term can be ignored. Up to the first order of approximation, the variance of $s_y^2$, and $t_1$ and bias and variances of $t_2$ (ignoring fpc term) are respectively given by

$$\operatorname{var}(s_y^2) = \frac{S_y^4}{n} \{\beta_2(y) - 1\} \tag{1.6}$$

$$\operatorname{var}(t_1) = \frac{S_y^4}{n} [\{\beta_2(y) - 1\} + \{\beta_2(x) - 1\}(1 - 2C)] \tag{1.7}$$

$$B(t_2) = \frac{S_y^2}{n} [\{\beta_2(x) - 1\}\theta(\theta - C)] \tag{1.8}$$

$$\operatorname{var}(t_2) = \frac{S_y^4}{n} [\{\beta_2(y) - 1\} + \theta\{\beta_2(x) - 1\}(\theta - 2C)] \tag{1.9}$$

where $\theta = \frac{S_x^2}{S_x^2 + \beta_2(x)}$; $\beta_2(y) = \frac{\mu_{40}}{\mu_{20}^2}$; $\beta_2(x) = \frac{\mu_{04}}{\mu_{02}^2}$; $h = \frac{\mu_{22}}{(\mu_{20} \cdot \mu_{02})}$; $C = \frac{(h-1)}{\beta_2(x) - 1}$ and

$$\mu_{rs} = \frac{1}{N} \sum_{i=1}^{N} (y_i - \bar{Y})^r (x_i - \bar{X})^s.$$

From (1.8), we see that the estimator $t_2$ suggested by Upadhyay and Singh (1999) is a biased estimator. In some application bias is disadvantageous. This led authors to suggest almost unbiased estimators of $S_y^2$.

## 2. A class of ratio-type estimators

Consider $t_{Ri} = s_y^2 \left( \dfrac{S_x^2 + \beta_2(x)}{s_x^2 + \beta_2(x)} \right)^i$ such that $t_{Ri} \in R$, for $i = 1,2,3$; where R denotes the set of all possible ratio-type estimators for estimating the population variance $S_y^2$. We define a class of ratio-type estimators for $S_y^2$ as –

$$t_r = \sum_{i=1}^{3} w_i t_{Ri} \in R, \tag{2.1}$$

where $\sum_{i=1}^{3} w_i = 1$ and $w_i$ are real numbers. $\tag{2.2}$

For simplicity we assume that the population size N is large enough so that the fpc terms are ignored. We write

$$s_y^2 = S_y^2(1+e_0), s_x^2 = S_x^2(1+e_1)$$

such that $E(e_0) = E(e_1) = 0$.

Noting that for large N, $\dfrac{1}{N} \cong 0$ and $\dfrac{n}{N} \cong 0$, and thus to the first degree of approximation,

$$E(e_0^2) = \frac{\beta_2(y)-1}{n}, \; E(e_1^2) = \frac{\beta_2(x)-1}{n}, \; E(e_0 e_1) = \frac{(h-1)}{n} = \frac{[\beta_2(x)-1]C}{n}.$$

Expressing (2.1) in terms of e's we have

$$t_r = S_y^2(1+e_0) \sum_{i=1}^{3} a_i (1+\theta e_1)^{-i} \tag{2.3}$$

Assume that $|\theta e_1| < 1$ so that $(1 + \theta e_1)^i$ is expandable. Thus expanding the right hand side of the above expression (2.3) and retaining terms up to second power of e's, we have

$$t_r = S_y^2 \left[ 1 + e_0 - \sum_{i=1}^{3} a_i i \left( \theta e_1 + \theta e_0 e_1 - \left( \frac{i+1}{2} \right) \theta^2 e_1^2 \right) \right]$$

or

$$t_r - S_y^2 = S_y^2 \left[ e_0 - \sum_{i=1}^{3} a_i i \left( \theta e_1 + \theta e_0 e_1 - \left( \frac{i+1}{2} \right) \theta^2 e_1^2 \right) \right] \qquad (2.4)$$

Taking expectation of both sides of (2.3) we get the bias of $t_r$, to the first degree of approximation, as

$$B(t_r) = \frac{S_y^2}{2n} \left[ \{\beta_2(x) - 1\} \sum_{i=1}^{3} i a_i \theta(\theta i - 2C + \theta) \right] \qquad (2.5)$$

Squaring both sides of (2.4), neglecting terms involving power of e's greater than two and then taking expectation of both sides, we get the mean-squared error of $t_r$ to the first degree of approximation, as

$$MSE(t_r) = \frac{S_y^4}{n} \left[ \{\beta_2(y) - 1\} + R_1 \{\theta \beta_2(x) - 1\} \{\theta R_1 - 2C\} \right] \qquad (2.6)$$

where $R_1 = \sum_{i=1}^{3} i \cdot w_i$ \qquad (2.7)

Minimizing the MSE of $t_r$ in (2.7) with respect to $R_1$ we get the optimum value of $R_1$ as

$$R_1 = \frac{C}{\theta} \qquad (2.8)$$

Thus the minimum MSE of $t_r$ is given by

$$\min. MSE(t_r) = \frac{S_y^4}{n} \left[ \{\beta_2(y) - 1\} - \{\beta_2(x) - 1\} C^2 \right]$$

$$= \frac{S_y^4}{n}\left[\{\beta_2(y)-1\}(1-\rho_1^2)\right] \qquad (2.9)$$

where $\rho_1 = \dfrac{(h-1)}{\sqrt{\{\beta_2(x)-1\}\{\beta_2(y)-1\}}}$ is the correlation coefficient between $(y-\bar{Y})^2$ and $(x-\bar{X})^2$.

From (2.2), (2.7) and (2.8) we have

$$\sum_{i=1}^{3} w_i = 1 \qquad (2.10)$$

and

$$\sum_{i=1}^{3} i w_i = \frac{C}{\theta} = \frac{\rho_1}{\theta}\left\{\frac{\beta_2(y)-1}{\beta_2(x)-1}\right\}^{\frac{1}{2}} \qquad (2.11)$$

From (2.10) and (2.11) we have three unknown to be determined from two equations only. It is therefore, not possible to find a unique value of the constants $w_i's(i=1,2,3)$. Thus in order to get the unique values of the constants $w_i's(i=1,2,3)$, we shall impose a linear constraint as

$$B(t_r) = 0 \qquad (2.12)$$

which follows from (2.5) that

$$(\theta-C)a_1 + (3\theta-2C)a_2 + (6\theta-3C)a_3 = 0 \qquad (2.13)$$

Equation (2.10), (2.11) and (2.13) can be written in the matrix form as

$$\begin{bmatrix} 1 & 1 & 1 \\ 1 & 2 & 3 \\ (\theta-C) & (3\theta-2C) & (6\theta-3C) \end{bmatrix} \begin{bmatrix} w_1 \\ w_2 \\ w_3 \end{bmatrix} = \begin{bmatrix} 1 \\ C/\theta \\ 0 \end{bmatrix} \qquad (2.14)$$

Using (2.14) we get the unique values of $w_i's(i=1,2,3)$ as

$$\left.\begin{aligned} w_1 &= \frac{1}{\theta^2}\left[3\theta^2 - 3\theta C + C^2\right] \\ w_2 &= \frac{1}{\theta^2}\left[-3\theta^2 + 5\theta C - 2C^2\right] \\ w_3 &= \frac{1}{\theta^2}\left[\theta^2 - 2\theta C + C^2\right] \end{aligned}\right\} \quad (2.15)$$

Use of these $w_i's (i=1,2,3)$ remove the bias up to terms of order $o(n^{-1})$ at (2.1). Substitution of (2.14) in (2.1) yields the almost unbiased optimum ratio-type estimator of the population variance $S_y^2$.

### 3. A class of product-type estimators

Consider $t_{Pi} = s_y^2 \left[\dfrac{s_x^2 + \beta_2(x)}{S_x^2 + \beta_2(x)}\right]^i$ such that $t_{Pi} \in P$, for $i=1,2,3$; where P denotes the set of all possible product-type estimators for estimating the population variance $S_y^2$.

We define a class of product-type estimators for $S_y^2$ as –

$$t_P = \sum_{i=1}^{3} k_i t_{Pi} \in P, \qquad (3.1)$$

where $k_i's\ (i=1,2,3)$ are suitably chosen scalars such that

$\sum\limits_{i=1}^{3} k_i = 1$ and $k_i$ are real numbers.

Proceeding as in previous section, we get

$$B(t_P) = \frac{S_y^2}{2n}\left[\{\beta_2(x) - 1\}\sum_{i=1}^{3} i a_i \theta(\theta i + 2C - \theta)\right] \qquad (3.2)$$

$$MSE(t_P) = \frac{S_y^4}{n} \left[ \{\beta_2(y) - 1\} + R_2 \theta \{(\beta_2(x) - 1)\}(\theta R_2 + 2C) \right] \quad (3.3)$$

where, $R_2 = \sum_{i=1}^{3} i k_i$ \quad (3.4)

Minimizing the MSE of $t_P$ in (3.4) with respect to $R_2$, we get the optimum value of $R_2$ as

$$R_2 = -\frac{C}{\theta} \quad (3.5)$$

Thus the minimum MSE of $t_P$ is given by

$$\min. MSE(t_P) = \frac{S_y^4}{n} \{\beta_2(y) - 1\}(1 - \rho_1^2) \quad (3.7)$$

which is same as that of minimum MSE of $t_r$ at (2.9).

Following the approach of previous section, we get

$$\left.\begin{aligned} k_1 &= \frac{1}{\theta^2} \left[ 3\theta^2 + 2\theta C + C^2 \right] \\ k_2 &= -\frac{1}{\theta^2} \left[ 3\theta^2 + 3\theta C + 2C^2 \right] \\ k_3 &= \frac{1}{\theta^2} \left[ \theta^2 + \theta C + C^2 \right] \end{aligned}\right\} \quad (3.8)$$

Use of these $k_i$'s (i=1,2,3) removes the bias up to terms of order O ($n^{-1}$) at (3.1).

4. **Empirical Study**

The data for the empirical study are taken from two natural population data sets considered by Das (1988) and Ahmed *et. al.* (2003).

**Population I** – Das (1988)

The variables and the required parameters are:

X: number of agricultural laborers for 1961.

Y: number of agricultural laborers for 1971.

$\beta_2(x) = 38.8898, \beta_2(y) = 25.8969, h=26.8142, S_x^2 = 1654.44$.

**Population II** – Ahmed et. al. (2003)

The variables and the required parameters are:

X: number of households

Y: number of literate persons

$\beta_2(x) = 8.05448, \beta_2(y) = 10.90334, S_x^2 = 11838.85, h=7.31399$.

In table 4.1 the values of scalars $w_i$'s (i=1,2,3) and $k_i$'s (i=1,2,3) are listed.

**Table 4.1:   Values of scalars $w_i$'s and $k_i$'s (i=1,2,3)**

| Scalars | Population | | Scalars | Population | |
|---|---|---|---|---|---|
| | I | II | | I | II |
| **w₁** | 1.3942 | 1.1154 | k₁ | 4.8811 | 5.5933 |
| **w₂** | -0.4858 | -0.1261 | k₂ | -6.0647 | -7.2910 |
| **w₃** | 0.0916 | 0.0109 | k₃ | 2.1837 | 2.6978 |

Using these values of $w_i$'s and $k_i$'s (i=1,2,3) given in table 4.1, one can reduce the bias to the order $O(n^{-1})$ respectively, in the estimators $t_r$ and $t_p$ at (2.1) and (3.1).

In table 4.2 percent relative efficiency (PRE) of $s_y^2, t_1, t_2, t_r$ (in optimum case) and $t_p$ (in optimum case) are computed with respect to $s_y^2$.

**Table 4.2: PRE of different estimators of $S_y^2$ with respect to $s_y^2$**

| Estimators | PRE $(., S_y^2)$ | |
|---|---|---|
| | **Population I** | **Population II** |
| $s_y^2$ | 100 | 100 |
| $t_1$ | 223.14 | 228.70 |
| $t_2$ | 235.19 | 228.76 |
| $t_r$ (optimum) | 305.66 | 232.90 |
| $t_p$ (optimum) | 305.66 | 232.90 |

Table 4.2 clearly shows that the suggested estimators $t_r$ and $t_p$ in their optimum case are better than the usual unbiased estimator $s_y^2$, Isaki's (1983) estimator $t_1$ and Upadhayaya and Singh (1999) estimator $t_2$.

**References**

Ahmed, M.S., Abu Dayyeh, W. and Hurairah, A. A. O. (2003): Some estimators for finite population variance under two-phase sampling. Statistics in Transition, 6, (1), 143-150.

Das, A.K. (1988): Contributions to the theory of sampling strategies based on auxiliary information. Ph.D thesis submitted to BCKV, Mohanpur, Nadia, and West Bengal, India.

Isaki, C. T. (1983): Variance estimation using auxiliary information. Journal of American Statistical Association.

Singh, S. and Singh, R. (1993): A new method: Almost Separation of bias precipitates in sample surveys. Journal of Indian Statistical Association, 31,99-105.

Upadhyaya, L.N. and Singh, H. P. (1999): An estimator for population variance that utilizes the kurtosis of an auxiliary variable in sample surveys. Vikram Mathematical Journal, 19, 14-17.

# A General Family of Estimators for Estimating Population Variance Using Known Value of Some Population Parameter(s)

Rajesh Singh, Pankaj Chauhan, Nirmala Sawan

School of Statistics, DAVV, Indore (M.P.), India

(rsinghstat@yahoo.com)

Florentin Smarandache

Department of Mathematics, University of New Mexico, Gallup, USA

(fsmarandache@yahoo.com)

**Abstract**

A general family of estimators for estimating the population variance of the variable under study, which make use of known value of certain population parameter(s), is proposed. Some well known estimators have been shown as particular member of this family. It has been shown that the suggested estimator is better than the usual unbiased estimator, Isaki's (1983) ratio estimator, Upadhyaya and Singh's (1999) estimator and Kadilar and Cingi (2006). An empirical study is carried out to illustrate the performance of the constructed estimator over others.

**Keywords:** Auxiliary information, variance estimator, bias, mean squared error.

## 1. Introduction

In manufacturing industries and pharmaceutical laboratories sometimes researchers are interested in the variation of their produce or yields (Ahmed et.al. (2003)).

Let $(U = U_1, U_2, \ldots, U_N)$ denote a population of N units from which a simple random sample without replacement (SRSWOR) of size n is drawn. Further let y and x denote the study and the auxiliary variables respectively.

Let $\overline{Y} = \frac{1}{N}\sum_{i=1}^{N} y_i$ and $S_y^2 = \frac{1}{N-1}\sum_{i=1}^{n}(y_i - \overline{Y})^2$ denotes respectively the unknown population mean and population variance of the study character y. Assume that population size N is very large so that the finite population correction term is ignored. It is established fact that in most of the survey situations, auxiliary information is available (or may be made to be available diverting some of the resources) in one form or the other. If used intelligibly, this information may yield estimators better than those in which no auxiliary information is used.

Assume that a simple random sample of size n is drawn without replacement. The usual unbiased estimator of $S_y^2$ is

$$s_y^2 = \frac{1}{n-1}\sum_{i=1}^{n}(y_i - \overline{y})^2 \tag{1.1}$$

where $\overline{y} = \frac{1}{n}\sum_{i=1}^{n} y_i$ is the sample mean of y.

When the population mean square $S_x^2 = \frac{1}{N-1}\sum_{i=1}^{N}(x_i - \overline{X})^2$ is known, Isaki (1983) proposed a ratio estimator for $S_y^2$ as

$$t_1 = \left(\frac{s_y^2}{s_x^2}\right) S_x^2 \tag{1.2}$$

where $s_x^2 = \frac{1}{n-1}\sum_{i=1}^{n}(x_i - \overline{x})^2$ is an unbiased estimator of $S_x^2$.

Several authors have used prior value of certain population parameter(s) to find more precise estimates. The use of prior value of coefficient of kurtosis in estimating the population variance of study character y was first made by Singh et. al. (1973). Kadilar and Cingi (2006) proposed modified ratio estimators for the population variance using different combinations of known values of coefficient of skewness and coefficient of variation.

In this paper, under SRSWOR, we have suggested a general family of estimators for estimating the population variance $S_y^2$. The expressions of bias and mean-squared error (MSE), up to the first order of approximation, have been obtained. Some well known estimators have been shown as particular member of this family.

## 2. The suggested family of estimators

Motivated by Khoshnevisan et. al. (2007), we propose following ratio-type estimators for the population variance as

$$t = s_y^2 \frac{\left(aS_x^2 - b\right)}{[\alpha(as_x^2 - b) + (1-\alpha)(aS_x^2 - b)]} \tag{2.1}$$

where $(a \neq 0)$, b are either real numbers or the function of the known parameters of the auxiliary variable x such as coefficient of variation C(x) and coefficient of kurtosis $(\beta_2(x))$.

The following scheme presents some of the important known estimators of the population variance, which can be obtained by suitable choice of constants $\alpha$, a and b:

**Table 2.1 : Some members of the proposed family of the estimators 't'**

| Estimator | Values of | | |
|---|---|---|---|
| | α | a | b |
| $t_0 = s_y^2$ | 0 | 0 | 0 |
| $t_1 = \dfrac{s_y^2}{s_x^2} S_x^2$  Isaki (1983) estimator | 1 | 1 | 0 |
| $t_2 = \dfrac{s_y^2}{s_x^2 - C_x}[S_x^2 - C_x]$ Kadilar and Cingi (2006) estimator | 1 | 1 | $C_x$ |
| $t_3 = \dfrac{s_y^2}{s_x^2 - \beta_2(x)}[S_x^2 - \beta_2(x)]$ | 1 | 1 | $\beta_2(x)$ |
| $t_4 = \dfrac{s_y^2}{s_x^2 \beta_2(x) - C_x}[S_x^2 \beta_2(x) - C_x]$ | 1 | $\beta_2(x)$ | $C_x$ |
| $t_5 = \dfrac{s_y^2}{s_x^2 C_x - \beta_2(x)}[S_x^2 C_x - \beta_2(x)]$ | 1 | $C_x$ | $\beta_2(x)$ |
| $t_6 = \dfrac{s_y^2}{s_x^2 + \beta_2(x)}[S_x^2 + \beta_2(x)]$ Upadhyaya and Singh (1999) | 1 | 1 | $-\beta_2(x)$ |

The MSE of proposed estimator 't' can be found by using the firs degree approximation in the Taylor series method defined by

$$MSE(t) \cong d \sum d' \qquad (2.2)$$

where

$$h = [\left.\frac{\partial h(a,b)}{\partial a}\right|_{S_y^2,S_x^2} \quad \left.\frac{\partial h(a,b)}{\partial b}\right|_{S_y^2,S_x^2}]$$

$$\Sigma = \begin{bmatrix} V(s_y^2) & Cov(s_y^2, s_x^2) \\ Cov(s_x^2, s_y^2) & V(s_x^2) \end{bmatrix}.$$

Here $h(a,b) = h(s_y^2, s_x^2) = t$. According to this definition, we obtain 'd' for the proposed estimator, t, as follows:

$$d = [\ 1 \quad -\frac{\alpha a S_y^2}{aS_x^2 + b}]$$

MSE of the proposed estimator t using (2.2) is given by

$$MSE(t) \cong V(s_y^2) - 2\alpha\left(\frac{aS_y^2}{aS_x^2 - b}\right)Cov(s_y^2, s_x^2) + \left(\frac{\alpha a S_y^2}{aS_x^2 - b}\right)V(s_x^2) \qquad (2.3)$$

where

$$\left.\begin{array}{l} V(s_y^2) = \lambda S_y^4[\beta_2(y) - 1] \\ V(s_x^2) = \lambda S_y^4[\beta_2(x) - 1] \\ Cov(s_y^2, s_x^2) = \lambda S_y^2 S_x^2 (h - 1) \end{array}\right\} \qquad (2.4)$$

where $\lambda = \frac{1}{n}$, $\beta_2(y) = \frac{\mu_{40}}{\mu_{20}^2}$, $\beta_2(x) = \frac{\mu_{04}}{\mu_{02}^2}$, $h = \frac{\mu_{22}}{\mu_{20}\mu_{02}}$,

$\mu_{rs} = \frac{1}{N}\sum_{i=1}^{N}(y_i - \overline{Y})^r (x_i - \overline{X})^s$, (r, s) being non negative integers.

Using (2.4), MSE of t can be written as

$$MSE(t) \cong \lambda S_y^4 \{\beta_2(y) - 1 - 2\alpha\theta(h - 1) + \alpha^2\theta^2(\beta_2(x) - 1)\} \qquad (2.5)$$

where $\theta = \frac{aS_x^2}{aS_x^2 - b}$.

The MSE equation of estimators listed in Table 2.1 can be written as-

$$\text{MSE}(t_i) \cong \lambda S_y^4 \{\beta_2(y) - 1 - 2\alpha\theta_i(h-1) + \alpha^2\theta_i^2(\beta_2(x) - 1)\}, \quad i = 2,3,4,5,6 \tag{2.6}$$

where

$$\theta_2 = \frac{S_x^2}{S_x^2 - C_x}, \qquad \theta_3 = \frac{S_x^2}{S_x^2 - \beta_2(x)},$$

$$\theta_4 = \frac{S_x^2\beta_2(x)}{S_x^2\beta_2(x) - C_x}, \quad \theta_5 = \frac{S_x^2 C_x}{S_x^2 C_x - \beta_2(x)}, \qquad \theta_6 = \frac{S_x^2}{S_x^2 + \beta_2(x)}.$$

Minimization of (2.5) with respect to $\alpha$ yields its optimum value as

$$\alpha = \frac{C}{\theta} = \alpha_{opt} \tag{2.7}$$

where $C = \frac{(h-1)}{\{\beta_2(x) - 1\}}$.

By substituting $\alpha_{opt}$ in place of $\alpha$ in (2.5) we get the resulting minimum variance of t as

$$\min.\text{MSE}(t) = \lambda S_y^4 [\beta_2(y) - 1 - \{\beta_2(x) - 1\}] \tag{2.8}$$

## 3. Efficiency comparisons

Up to the first order of approximation, variance (ignoring finite population correction) of $t_o = s_y^2$ and $t_1$ is given by –

$$\text{Var}(s_y^2) = \lambda S_y^4 [\beta_2(y) - 1] \tag{3.1}$$

$$\text{MSE}(t_1) = \lambda S_y^4 [\{\beta_2(y) - 1\} + \{\beta_2(x) - 1\}(1 - 2C)] \tag{3.2}$$

From (2.6), (2.8), (3.1), and (3.2), we have

$$\text{Var}(s_y^2) - \min.\text{MSE}(t) = \lambda S_y^4 \{\beta_2(x) - 1\} C^2 > 0 \tag{3.3}$$

$$\text{MSE}(t_i) - \min.\text{MSE}(t) = \lambda S_y^4 \{\beta_2(x) - 1\}(\theta_i - C^2) > 0, \quad i = 1,2,3,4,5,6 \tag{3.4}$$

provided $C \neq \theta_i$.

Thus it follows from (3.3) and (3.4) that the suggested estimator under 'optimum' condition is (i) always better then $s_y^2$, (ii) better than Isaki's (1983) estimator $t_1$ except when C = 1 in which both are equally efficient, and (iii) Kadilar and Cingi (2006) estimators $t_i (i = 2,3,4,5)$ except when $C = \theta_i (i = 2,3,4,5)$ in which t and $t_i (i = 2,3,4,5)$ are equally efficient.

## 4. Empirical study

We use data in Kadilar and Cingi (2004) to compare efficiencies between the traditional and proposed estimators in the simple random sampling.

In Table 4.1, we observe the statistics about the population.

**Table 4.1: Data statistics of the population for the simple random sampling**

N = 106, n = 20, $\rho = 0.82$, $C_y = 4.18$, $C_x = 2.02$, $\overline{Y} = 15.37$, $\overline{X} = 243.76$, $S_y = 64.25$, $S_x = 491.89$, $\beta_2(x) = 25.71$, $\beta_2(y) = 80.13$, $\lambda = 0.05$, $\theta = 33.30$.

The percent relative efficiencies of the estimators $s_y^2$, $t_i (i = 2,3,4,5,6)$ and $\min.MSE(t)$ with respect to $s_y^2$ have been computed and presented in Table 4.2 below.

**Table 4.2: Relative efficiencies (%) of** $s_y^2$**,** $t_i (i = 2,3,4,5,6)$ **and** $\min.MSE(t)$ **with respect to** $s_y^2$**.**

| Estimator | PRE $(., s_y^2)$ |
|---|---|
| $t_0 = s_y^2$ | 100 |
| $t_1$ | 201.6564 |
| $t_2$ | 201.6582 |
| $t_3$ | 201.6782 |
| $t_4$ | 201.6565 |
| $t_5$ | 201.6672 |
| $t_6$ | 201.6347 |
| $\min.MSE(t)$ | 214.3942 |

## 5. Conclusion

From theoretical discussion in section 3 and results of the numerical example, we infer that the proposed estimator 't' under optimum condition performs better than usual estimator $s_y^2$, Isaki's (1983) estimator $t_1$, Kadilar and Cingi's (2006) estimators ($t_2$, $t_3$, $t_4$, $t_5$) and Upadhyaya and Singh's (1999) estimator $t_6$.

## References

Ahmed, M.S., Dayyeh, W.A. and Hurairah, A.A.O. (2003): Some estimators for finite population variance under two-phase sampling. Statistics in Transition, 6, 1, 143-150.

Isaki, C.T. (1983): Variance estimation using auxiliary information. Jour. Amer. Stat. Assoc., 78, 117-123.

Kadilar, C. and Cingi, H. (2006): Ratio estimators for the population variance in simple and stratified random sampling. Applied Mathematics and Computation 173 (2006) 1047-1059.

Singh, J., Pandey, B.N. and Hirano, K. (1973): On the utilization of a known coefficient of kurtosis in the estimation procedure of variance. Ann. Inst. Statist. Math., 25, 51-55.

Upadhyaya, L.N. and Singh, H. P. (1999): An estimator for population variance that utilizes the kurtosis of an auxiliary variable in sample surveys. Vikram Mathematical Journal, 19, 14-17.

**This volume is a collection of six papers on the use of auxiliary information and *a priori* values in construction of improved estimators. The work included here will be of immense application for researchers and students who employ auxiliary information in any form.**